\documentclass{scrartcl}

\usepackage[sumlimits,intlimits,namelimits]{amsmath}
\usepackage{amssymb,pstricks,pst-plot}
\usepackage[english]{babel}
\usepackage{bbm}
\usepackage{parskip}
\usepackage{slashed}

\newcommand{\eqnl}[2]{\begin{eqnarray}#1\label{#2}\end{eqnarray}}
\newcommand{\eqnn}[1]{
\begin{eqnarray*}#1\end{eqnarray*}}

\newcommand{\eqngrl}[3]{
\begin{eqnarray}#1\nonumber\\#2\label{#3}\end{eqnarray}}

\newcommand{\eqngrrl}[4]{
\begin{eqnarray}#1\nonumber\\#2\label{#4}\\
#3\nonumber\end{eqnarray}}

\newcommand{\refs}[1]{(\ref{#1})}
\newcommand{\ft}[2]{{\textstyle\frac{#1}{#2}}}

\newcommand{\mD}{\mathcal{D}}
\newcommand{\mH}{\mathcal{H}}
\newcommand{\mL}{\mathcal{L}}

\newcommand{\mN}{\mathcal{N}}
\newcommand{\mP}{\mathcal{P}}
\newcommand{\mZ}{\mathcal{Z}}
\newcommand{\mh}{\mathfrak{h}}

\newcommand{\uS}{\text{S}}
\newcommand{\uA}{\text{A}}

\newcommand{\ud}{\text{d}}
\newcommand{\ui}{\text{i}}
\newcommand{\ue}{\text{e}}
\newcommand{\uB}{\text{B}}
\newcommand{\uF}{\text{F}}

\newcommand{\tr}{\text{tr}}

\newcommand{\sign}{\text{sign}\;}
\renewcommand{\Im}{\text{Im}\,}
\renewcommand{\Re}{\text{Re}\,}

\newcommand{\N}{\mathbbm{N}}
\newcommand{\R}{\mathbbm{R}}
\newcommand{\C}{\mathbbm{C}}

\newcommand{\id}{\mathbbm{1}}

\def\up{\,\vert \!\uparrow\rangle}
\def\down{\,\vert \!\downarrow\rangle}
\def\vac{\vert 0\rangle}
\def\psid{\psi^\dagger}
\def\la{\langle}
\def\ra{\rangle}
\def\psib{\bar\psi}
\def\epsb{\bar\eps}
\def\Aab{A^a_{\hskip4pt b}}

\def\Wa{W,_a}
\def\Wab{W,_{ab}}
\def\ps{\!+\!}
\def\ms{\!-\!}
\def\mtxt#1{\quad\hbox{{#1}}\quad}
\def\eps{\epsilon}
\def\al{\alpha}
\def\gam{\gamma}
\def\lam{\lambda}
\newcommand{\ha}{\frac{1}{2}}

\newcommand{\h}{\ft12}

\newcommand{\ih}{\ft{\ui}{2}}
\newcommand{\p}{\partial}
\newcommand{\pL}{\partial_\text{\tiny L}}

\newcommand{\pR}{\partial_\text{\tiny R}}
\newcommand{\pS}{\partial_\text{\tiny S}}
\newcommand{\imR}{\hat{p}_\text{\tiny R}}
\newcommand{\imL}{\hat{p}_\text{\tiny L}}
\newcommand{\imRL}{\hat{p}_\text{\tiny R+L}}

\newcommand{\pRL}{\partial_\text{\tiny R+L}}
\newcommand{\pSL}{\partial_\text{\tiny SLAC}}
\newcommand{\imSL}{\hat{p}_\text{\tiny SLAC}}
\newcommand{\pA}{\partial_\text{\tiny A}}
\newcommand{\hF}{h_\text{\tiny F}}
\newcommand{\HB}{H_\text{\tiny B}}
\newcommand{\HF}{H_\text{\tiny F}}
\newcommand{\PsiB}{\Psi_\text{\tiny B}}
\newcommand{\PsiF}{\Psi_\text{\tiny F}}
\newcommand{\EB}{E_\text{\tiny B}}
\newcommand{\EF}{E_\text{\tiny F}}
\newcommand{\va}{|0\rangle}
\newcommand{\di}{\slashed{D}}
\newcommand{\fdi}{\slashed{\partial}\hskip1pt}

\begin{document}
\setlength{\parindent}{0cm}


\title {From the Dirac Operator to Wess-Zumino Models
on Spatial Lattices}
\author{A. Kirchberg, J.D. L\"ange and A.
Wipf\footnote{\tt{A.Kirchberg@tpi.uni-jena.de,
J.D.Laenge@tpi.uni-jena.de, A.Wipf@tpi.uni-jena.de}} \\[10pt]
Theoretisch--Physikalisches Institut, \\
Friedrich--Schiller--Universit\"at Jena,\\
07743 Jena, Germany}
\date{}


\maketitle

\vspace{-250pt}
\hfill{FSU-TPI 04/04}
\vspace{+250pt}

\paragraph{Abstract:}
We investigate two-dimensional Wess-Zumino models
in the continuum and on spatial lattices in
detail. We show that a non-antisymmetric
lattice derivative not only excludes chiral
fermions but in addition introduces supersymmetry
breaking lattice artifacts. We study the
nonlocal and antisymmetric SLAC derivative
which allows for chiral fermions
without doublers and minimizes those
artifacts. The supercharges of the
lattice Wess-Zumino models are obtained by dimensional
reduction of Dirac operators in high-dimensional spaces.
The normalizable zero modes of the models with
$\mN=1$ and $\mN=2$ supersymmetry are counted
and constructed in the weak- and strong-coupling
limits. Together with known methods from operator
theory this gives us complete control of the
zero mode sector of these theories for arbitrary
coupling.



\clearpage


\clearpage

\section{Introduction}

Ever since its invention supersymmetry has been
an important subject in high-energy physics beyond
the standard model. It is considered to be a necessary ingredient
to bridge the gap between the scale of electroweak symmetry
breaking and the much larger unification scale.
Nowadays, supersymmetric theories cover the whole range
from supersymmetric classical mechanics
\cite{Galvao:1980cu},
quantum mechanics
\cite{Nicolai:1976xp, Witten:1981nf},
scalar and gauge field theories
\cite{Sohnius:1985qm}
to string- and $M$-theory
\cite{Green:1987sp}.
They allow for the construction of low-energy effective
actions, as for the $\mN=2$ Seiberg-Witten model
\cite{Seiberg:1994rs} or the
formulation of certain duality
relations, like in the original Maldacena
conjecture for gauge theories with
$\mN=4$ extended supersymmetry \cite{Maldacena:1998re}.

The non-perturbative effects in supersymmetric theories,
and in particular, the dynamical breaking of supersymmetry
are a subject of intensive studies. At present time
the lattice formulation is the only tool for
\emph{systematic} investigations of such effects, and lattice
simulations provide the means of doing reliable calculations
in the strong-coupling regime or near a phase
transition point. After the pioneering work of
Dondi and Nicolai \cite{Dondi:1977tx} there
has been an ongoing effort into formulating,
understanding and simulating supersymmetric
theories on the lattice
\cite{Elitzur:1982vh,Banks:1982ut,Bartels:1983wm,Montvay:1998ak}.
Recent lattice results, e.g. on the breaking of
supersymmetry, have been obtained in
\cite{Catterall:2003ae,Montvay:2001aj,Farchioni:2001wx}.

A commonly accepted guiding principle in any good lattice calculation
is to build in as many of the symmetries of the continuum model as
possible, such that the lattice results respect these symmetries
identically. However, often these are conflicting requirements and not
all symmetries can be incorporated on the lattice. This in turn
introduces subtle lattice artifacts into the formalism, which
one may not get rid of in the continuum limit.
For example, lattice regularizations of supersymmetric theories generically
\emph{break large parts of supersymmetry},
and it is a nontrivial problem to recover supersymmetry in the
continuum limit. However, there are discretizations with highly
nonlocal derivative operators, for which supersymmetry is
manifestly realized \cite{Dondi:1977tx,Bartels:1983wm}.
Alternatively, for two-dimensional models one can
discretize only space (time remains continuous) such that a
subalgebra of the $\mN=1$ supersymmetry algebra,
\eqnn{
\{Q_\al,Q_\beta\}=
2(\gam^\mu\gam^0)_{\al\beta}P_\mu}
remains intact
\cite{Elitzur:1982vh,Banks:1982ut,Rittenberg:1982gd}.
That subalgebra then determines the spectral properties of the
super-Hamiltonian $H$.
The \emph{fermion doubling} for naive lattice derivatives
\cite{Nielsen:1981rz,Friedan:1982nk}
is another apparently unrelated notorious example of
such lattice artifacts. For bosons there is no such problem.
However, if we try to preserve part of
supersymmetry then the fermionic mirror states
lead to doublers in the bosonic sector as well.

In this paper we study continuum and lattice
versions of two-dimensional Wess-Zumino (WZ) models.
Similar to the original four-dimensional theory \cite{Wess:1974tw},
these models contain scalar and
fermion fields coupled by a Yukawa
term. A particular version possesses
$\mN=2$ supersymmetry and has been the subject
of analytic \cite{Nicolai:1980nr,Elitzur:1983nj}  and numerical
\cite{Catterall:2001fr} studies.

In section $2$ we consider the off-shell formulation
for a general class of continuum models and derive
the supersymmetry transformations and Noether currents.
Particular emphasis is put on the form of the
central charges \cite{Witten:1978mh}.

In section $3$ we turn to the lattice version
of the models. We show that for real and \emph{antisymmetric}
lattice derivatives the $\mN=1$ algebra can be represented
on free fields. The local left- and right-derivatives
are not antisymmetric and the anticommutator of
the corresponding supercharges does not yield
the discretized Hamiltonian for the free model.
If we insist that supersymmetry is realised on free
fields without fermion and boson doubling then
we must allow for nonlocal derivatives on the lattice.
One particular such derivative, the SLAC operator, is
introduced in this section. The numerical results for this
operator concerning supersymmetry in lower-dimensional
systems are in excellent agreement with continuum results. 
In section $4$ we show how to derive the models with
$\mN=1$ and $\mN=2$ supersymmetry on a spatial lattice
by a suitable \emph{dimensional reduction} of a high-dimensional
Euclidean Dirac operator. In the process of reduction
the Dirac matrices and coordinates turn into
Majorana spinors and scalar fields on the lattice.
We count and \emph{construct} the normalisable
eigenstates of $H$ with zero energy both
in the weak and strong-coupling limits. In particular
we find that the $\mN=2$ models with $\phi^{2q}$
interaction admit $q^N$ such states
if $N$ is the number of spatial lattice sites.

In section $4$ we bridge the gap between strong- and
weak-coupling regimes for models with $\mN=1$ and
$\mN=2$ supersymmetry with the help of powerful methods
from operator theory. Using a theorem
by Kato we prove that the zero modes in the
strong-coupling limit survive for intermediate
couplings as long as the coupling constant of
the leading term in the potential does not vanish.
We comment on what we expect to happen in the
continuum limit of the $\mN=2$ models, where only
$q$ of the $q^N$ zero modes survive \cite{Jaffe:1987qz}.
We also comment on recent lattice simulations of
the two dimensional Wess-Zumino model by
Beccaria et al. \cite{Beccaria:2004pa}.
Some technical details concerning the nonlocal
SLAC operator and the proof that the transition
from strong to intermediate couplings
is governed by a relative compact perturbation are relegated
to the appendix.


\section{Wess-Zumino Models in $1+1$ Dimensions}
\label{chapwzm}
In the off-shell formulation
two-dimensional parity invariant Wess-Zumino models
contain a set of, say $d$, triples, each containing a
real scalar $\phi$, Majorana spinor $\psi$ and auxiliary
field $F$. In a Majorana representation for the Clifford algebra
\eqnl{
\{ \gam^\mu , \gam^\nu \} = 2 \eta^{\mu\nu}, \mtxt{with}
\gam^0 {\gam^\mu}^\dagger \gam^0 = \gam^\mu,\quad
\eta=\hbox{diag}(1,-1),}{algebra1}
the Majorana spinors are real.

The supersymmetry algebra is spanned by $\mN$ Hermitian
spinorial supercharges $Q^{(I)}$, $I=1,\ldots,\mN$,
by the Hermitian two-momentum $P_\mu$ and by the (anti-)symmetric matrix
of Hermitian central charges $Z^{IJ}_\uS$ ($Z^{IJ}_\uA$)
and has the form
\eqnl{
\{ Q_\al^{(I)} , \bar{Q}_\beta^{(J)} \} = 2 \left( \delta^{IJ}
\slashed{P}_{\al\beta} + \ui \delta_{\al\beta}
\mZ^{IJ}_\uA + \ui \gam_{*\al \beta}
\mZ^{IJ}_\uS \right),\quad \gam_*=\gam^0\gam^1,}{algebra2}
with spinor index $\al=1,2$.

In component fields the Lagrangian of the
models with $\mN=1$ supersymmetry reads \cite{Alvarez-Gaume:1983ab}
\eqnl{
\mL = \h \p_\mu \phi^a \p^\mu \phi_a -
F^a \Wa +\h F^a F_a +
\ih \psib^a \fdi\psi_a -\h \Wab \psib^a
\psi^b,}{wess2}
where the \emph{superpotential} $W$ depends on
the dimensionless scalar fields $\phi^1,\dots,\phi^d$.
We denoted the derivative of $W$ with respect to $\phi^a$
by $W,_a$ and employed the Einstein
summation convention. For Wess-Zumino models
the target spaces are $\R^d$ with Euclidean metric $\delta_{ab}$.

Now we consider the most general \emph{linear} off-shell supersymmetry
transformation of the fields. Since
$(\phi^a,\psi^a,F^a)$ have mass dimensions $(0,\ft12,1)$
respectively, such transformations have the form  \cite{Alvarez-Gaume:1983ab}
\eqngrrl{
\delta_\eps\phi^a  & = & \epsb(A\psi)^a,}
{\delta_\eps\psi^a & = &\ui \fdi (B\phi)^a\eps+(CF)^a\eps,}
{\delta_\eps F^a   & = &\ui \epsb\fdi(D\psi)^a,}{wess4}
where, for example, $(A\psi)^a= \Aab\psi^b$.
The constant matrices $A,B,C,D$ must be real
for the supersymmetry variations to be Hermitian fields.
The requirement that $\mL$ transforms into
a divergence implies the following algebraic relations
for these matrices and the real symmetric matrix
$W''=(W,_{ab})$,
\begin{eqnarray}
A+B^T & = & 0, \quad D+C^T=0,\label{wess5}\\
A^T W''+ W''C & = & 0, \quad W'' A^T +C W''=0.\label{wess6}
\end{eqnarray}
It follows that
\eqnn{\delta \mL=
\epsb\, \p_\mu V^\mu+\Delta\mtxt{with}
\Delta=-\ft12 W,_{abc}\big(\epsb A^a_{\hskip 3pt d}\psi^d\big)
\;\big(\psib^b\psi^c\big).}
Free models have quadratic
superpotentials and $\Delta$ is identically zero.
For interacting models we may exploit the
Fierz identity
\eqnn{
(\psib^a\psi^b)(\psi^c\psi^d)
+(\psib^a\psi^d)(\psi^b\psi^c)+(\psib^a\psi^c)(\psi^d\psi^b)=0}
to prove that $\Delta$ vanishes, provided
\eqnl{
W''A=A^TW''}{wess8}
holds true. Then the action is left invariant
by the transformations \refs{wess4} and the
corresponding conserved Noether current reads
\eqnl{
J^\mu=
\left(\p^\mu\phi-\gam_*\eps^{\mu\nu}\p_\nu \phi\right)^a
(A\psi)^a-\ui(CW')^a \gam^\mu \psi^a,\quad
W'=\left(\partial W/\partial \phi^a\right).}{wess9}
In what follows, employing \refs{wess5}, we express the matrices 
$B$ and $D$ in terms of $A$ and $C$. We consider $\mN$ supersymmetries
\refs{wess4} with matrices $(A_I,C_I)$
and denote the corresponding supersymmetry transformations
by $\delta^{(I)}_\eps$.
For all pairs $(A_I,C_I)$ the conditions
\refs{wess6} and \refs{wess8} must hold
for the Lagrangian to be invariant.
These conditions severely restrict the form of the
superpotential $W$. We also demand that two
supersymmetry transformations close on translations (later
we shall comment on the possibility of central charges)
\eqnl{
\big[\,\delta^{(I)}_{\eps_1},\delta^{(J)}_{\eps_2}
\big]\Phi=
2\ui\delta^{IJ}(\epsb_2\gam^\mu\eps_1)\p_\mu\Phi,}{wess10}
and this puts further restrictions on the matrices.
For the scalar and the auxiliary field the condition
\refs{wess10} read
\eqnl{
A_IA^T_J+A_JA^T_I=C^T_IC_J+C_J^TC_I=2\delta_{IJ}\id\mtxt{and}
A_I C_J-A_JC_I=0.}{wess11}
In particular all matrices are orthogonal,
such that the two conditions in \refs{wess6} coincide.
Actually, the last relation implies that the algebra \refs{wess10}
is realized on the Majorana fields as well.

The transformation $\delta^{(I)}_\eps$ is generated
by the Noether charge corresponding to $J^\mu_I$ in \refs{wess9},
\eqnl{
Q^{(I)}=
\int \ud x \left((\pi-\phi'\gam_*)_a(A_I\psi)^a
-\ui(C_IW')_a \gam^0 \psi^a\right),\quad \pi^a=\dot\phi^a,}{wess12}
where we have set $(\ud \phi^a/\ud x)=\phi'$.

\textbf{Canonical structure:}
The canonical structure is more transparent
in the on-shell formulation. This is obtained
from the off-shell one by replacing $F_a$
by $W,_a$.
The nontrivial equal time (anti)commutators between
the scalar fields, their conjugated
momentum fields $\pi^a=\dot\phi^a$
and the Majorana fields read
\eqnl{
\{ \psi^a_\al(x),\psi^b_\beta(y) \}=\delta_{\al \beta}\delta^{ab}
\delta(x-y)\mtxt{and}
[\phi^a(x) , \pi^b(y) ]=\ui \delta^{ab} \delta(x-y).}{wess15}
The Hamiltonian is the Legendre transform of the Lagrangian,
\eqnl{
H=\int \ud x\,\mH,\qquad
\mH=
\h\pi\cdot\pi +\h\phi'\cdot\phi' + \h W'\cdot W'
+\h\psi^\dagger \hF\psi,}{wess17}
where, for example, $\pi\cdot\pi=\pi_a\pi^a$. We
have introduced the \emph{Hermitian} Dirac-Hamiltonian
\eqnl{
(\hF)_{ab}=-\ui\gam_*\p_x\delta_{ab}+\gam^0W,_{ab}
\equiv (\hF^0)_{ab}+\gam^0 W,_{ab}.}{wess20}
The action is invariant under spacetime translations
generated by Noether charges
\eqnl{
P_0=\mH\mtxt{and}
P_1=\int \ud x \; \left( \pi\cdot\phi' + \ih
\psib\gam^0\psi' \right),}{wess22}
and under supersymmetry transformations
\refs{wess4} generated by the above supercharges $Q^{(I)}$.
By using the relations (\ref{wess6},\ref{wess11})
one proves that the $Q^{(I)}$
satisfy the super-algebra \refs{algebra2} with central
charges
\eqnl{
\mZ^{IJ}_A=0\mtxt{and}
\mZ^{IJ}_S=
-\int \phi'\cdot \big(A_IC_J\big)W',}{wess24}
where we have neglected ambiguous surface terms
containing the Majorana fields only.
Note, that the integrand
is a total derivative, since the \emph{integrability
conditions} for the existence of a potential
$U(\phi(x))$ with
\eqnn{
\phi'\cdot(A_IC_J)W'=\frac{\ud U}{\ud x}=U'\cdot\phi'}
is that $A_IC_JW''$ is a symmetric matrix.
But this follows from the condition \refs{wess6}.  

In most explicit calculations we choose the Majorana
representation
\eqnl{
\gam^0=\sigma_2,\quad \gam^1=\ui\sigma_3
\mtxt{and}
\gam_* = \gam^0 \gam^1 = - \sigma_1}{wess26}
such that the superalgebra takes the simple form
\eqngrrl{
\{ Q_1^{(I)} , Q_1^{(J)} \} &=& 2\left(H\delta^{IJ}+\mZ_S^{IJ}\right),}
{\{ Q_2^{(I)} , Q_2^{(J)} \} &=& 2\left(H\delta^{IJ}-\mZ_S^{IJ}\right),}
{\{ Q_1^{(I)} , Q_2^{(J)} \}&=& 2 \left(P_1\delta^{IJ}+\mZ_A^{IJ}\right).}
{wess27}

\textbf{$\mN=1$ supersymmetry:}
There is always at least one solution to the constraints
(\ref{wess5},\ref{wess6},\ref{wess8}) and \refs{wess11}
for an arbitrary superpotential $W$, namely
\eqnl{
A_1=-B_1=-C_1= D_1=\id.}{wess28}
Solving for the auxiliary field, $F_a= W,_a$,
the on-shell transformations take the form
\eqnl{
\delta_\eps^{(1)}\phi= \bar{\eps} \psi,
\quad
\delta_\eps^{(1)}\psi =\left(-\ui \fdi\phi-W'\right)\eps,}{wess30}
and the corresponding supercharge reads
\eqnl{
Q^{(1)}=\int \ud x\; \left( \pi -\phi' \gam_*
+ \ui W' \gam^0\right)\cdot\psi\;.}{wess32}
For vanishing spinors the only non-trivial central charge is 
\eqnl{
\mZ_S=\int \ud x\,\frac{\ud W}{\ud x}.}{wess33}

\textbf{$\mN=2$ extended supersymmetry:}
We assume that the model \refs{wess2}
admits a second supersymmetry besides the solution
\refs{wess28}. The conditions
\refs{wess11} imply
\eqnl{
A_2=-C_2=I,\quad I=-I^T,\quad I^2=-\id.}{wess34}
The matrix $I$ defines a complex structure
and exists for all target spaces $\R^d$ with even
dimension $d$. The conditions in \refs{wess6} and \refs{wess8}
on the superpotential both reduce to
\eqnl{
IW''+W''I=0,}{wess37}
which means that the superpotential is a
harmonic function of the scalar fields,
in agreement with the general analysis in
\cite{Alvarez-Gaume:1983ab}.
On-shell, the second supersymmetry has the form
\eqnl{
\delta^{(2)}_\eps\phi=\bar{\eps} I\psi,\quad
\delta^{(2)}_\eps\psi=\left(\ui \fdi I\phi-
IW'\right)\eps,}{wess38}
and is generated by the Noether-supercharge
\eqnl{
Q^{(2)} = \int \ud x\; \left(\pi - \phi' \gam_* - \ui W'
\gam^0 \right)\cdot(I\psi).}{wess40}
For vanishing spinor fields the central charges read
\eqnl{
\mZ_A^{IJ}=0\mtxt{and}
\left(\mZ_S^{IJ}\right)=\sigma_3\int \ud x\; \frac{\ud W}{\ud x}
-\sigma_1\int \ud x\;\frac{\ud U}{\ud x},}{wess42}
where $U$ is the imaginary part of
the analytic function $F(\phi^1 \ps\ui \phi^2)=W+\ui U$
with real part $W$.
\par
For the models with $\mN= 2$ supersymmetry
there exists a concise formulation in which
two real scalars are combined to a
complex scalar, and two Majorana spinors are
combined to a Dirac spinor. For example, for the
target space $\R^2$ we set
\eqnl{
\phi=\frac{1}{\sqrt{2}}\left(\phi^1+\ui\phi^2\right),\quad
\psi=\frac{1}{\sqrt{2}}\left(\psi^1+\ui\gam_*\psi^2\right).}{cf1}
The harmonic superpotential is the real part
of a holomorphic function,
\eqnl{
W(\phi,\bar\phi)=F(\phi)+\bar F(\bar\phi),}{cf3}
and the on-shell Lagrangian
takes the form
\eqnl{
\mL=\p_\mu\phi\p^\mu\phi^\dagger+\ui\bar\psi\fdi\psi
-\ft12 \vert F'\vert^2-F''\psib P_+\psi
-\bar F'' \psib P_-\psi,}{cf5}
where $F'$ is the derivative of $F$
with respect to the \emph{complex} field $\phi$ and
we have introduced the chiral projectors
\eqnl{
P_\pm=\h (\id+\gam_*).}{cf7}
Along with the real scalar fields one combines the
corresponding conjugate momentum fields
to a complex momentum, $\pi=(\pi^1\ms\ui\pi^2)/\sqrt{2}$,
such that
\eqnl{
[\phi(x),\pi(y)]=\ui \delta(x-y)\mtxt{and}
\{\psi_\al,\psi^\dagger_\beta\}=\delta_{\al\beta}.}{cf9}
The complex supercharge takes the form
\eqnl{
Q=\h\left(Q^{(1)}+\ui\gam_* Q^{(2)}\right)=
\left(\pi-\bar\phi'+\ui F'\gam^0\right)P_+\psi
+\left(\bar\pi+\phi'+\ui \bar F'\gam^0\right)P_-\psi.}{cf11}
and satisfies the anticommutation relations
\eqnl{
\{Q,Q\}=0\mtxt{and}
\{Q,\bar Q\}=\slashed{P}+\gam_* \mZ^{11}_S-\mZ^{12}_S.}{cf13}
\textbf{Higher supersymmetries:}
Next we show that with the absence of central
charges there is no third linear off-shell
supersymmetry besides \refs{wess30}
and \refs{wess38}. To be compatible
with the first transformation in \refs{wess30},
the orthogonal matrices $A_3$ and $C_3$ 
must be antisymmetric and of opposite sign.
The conditions \refs{wess11} between the
second and third supersymmetry imply
\eqnn{
[I,A_3]=\{I,A_3\}=0,}
which is impossible for orthogonal matrices
$I$ and $A_3$. We conclude that the models
\refs{wess2} admit at most two
linear off-shell supersymmetries.

Let us mention that, if we allow for central charges
in the superalgebra, there exist further supersymmetries.
But the corresponding models are massive free
models. They can be derived by a dimensional reduction
of the free $\mN=2$ model in $4$ dimensions.

\section{Lattice Formulations of Wess-Zumino Models}

As ultraviolet-cutoff we discretize space, introduce
a spatial lattice with $N$ equidistant sites
and choose periodic boundary conditions.
The time is kept continuous
such that time translations remain symmetries
generated by the Hamiltonian. Following
\cite{Elitzur:1982vh} we try to preserve
at least that subalgebra of \refs{algebra2}
which involves $H$.

The fields of the supersymmetric model in the Hamiltonian
formulation are discretized as follows, 
\eqnl{
\left(\phi^a(x), \pi^a(x), \psi^a(x)\right)
\longrightarrow \left(\phi^a(n), \pi^a(n), \psi^a(n)\right),\quad n=1,
\ldots,N,}{lmod1}
where the lattice spacing has been set to one.
On a space-lattice the derivative becomes a
difference operator the particular choice of which
is left open for the moment being. We \emph{define} the
lattice Hamiltonian as square of the discretized
supercharge $Q_1$.
For interacting theories it consists of the
discretized Hamiltonian of the continuum theory
plus a lattice counterpart of the central charge.

On-shell the $\mN=1$ model contains $d\in\{1,2,\dots\}$
Hermitian scalar fields $\phi^a(n)$ and $d$
Majorana spinors $\psi^a(n)$ on $N$
lattice sites ($n=1, \ldots ,N$).
The fields obey the non-trivial
canonical (anti-)commutation relations
\eqnl{
\,[ \phi^a(n) , \pi^b(n') ]=\ui \delta^{ab}\delta(n,n')
\mtxt{and}
\{ \psi^a_\al(n) , \psi^b_\beta(n') \}=\delta^{ab}\delta_{\al \beta}
\delta(n,n').}{lmod3}
We choose a Majorana representation
such that the $\psi^a$ are Hermitian two
component spinors.

When we put the supercharge on a space-lattice, we must
choose the lattice derivative in the term
\eqnn{
-\int\phi'\gam_*\psi =\int\phi\gam_*\psi'=
\ui \int \phi \hF^0\psi}
in \refs{wess12}.
Since we do not want to specify $\p$ at this
point we make the general ansatz for the
Hermitian Dirac-Hamiltonian
\eqnl{
\hF^0=\ui\delta_{ab}\begin{pmatrix}0&\p\cr -\p^\dagger&0\end{pmatrix}
\;,\mtxt{with}\p\p^\dagger=\p^\dagger\p\equiv -\triangle}{lmod12}
and a real $\p$ with correct continuum
limit. $\p$ must be real, since it should map Majorana spinors into
Majorana spinors. Let us define its symmetric and
antisymmetric parts
\eqnl{
\pS=\h(\p+\p^\dagger),\quad
\pA=\h(\p-\p^\dagger)\mtxt{with}[\pA,\pS]=0,\quad
\pA^2-\pS^2=\triangle.}{lmod13}
The last two properties follow from our assumption
$[\p,\p^\dagger]=0$ in \refs{lmod12}.
Since
\eqnl{
\hF^0\stackrel{\refs{wess26}}{=}
-\ui\gam_*\pA-\gam^0\pS,}{lmod13a}
\emph{chirality is preserved} for massless fermions if
$\p=\pA$ is antisymmetric, in which case
$\hF^0=-\ui\gam_*\pA$. Thus, if $\p$ is antisymmetric and
local then, according to some long-standing
no-go theorems there is fermion doubling.
There are many such theorems, and
we mention only two, one due to Nielsen and
Ninomiya \cite{Nielsen:1981rz}
and a later elaboration due to Friedan \cite{Friedan:1982nk}.
No-go theorems are notorious in that
people find a way around them, and following
Friedans work, L\"uscher \cite{Luscher:1998pq}
and others did so.
Below we circumvent the no-go theorems by using
a nonlocal and antisymmetric derivative.

However, most lattice derivative are not antisymmetric
in which case $\hF^0$ contains a momentum dependent
mass term $-\gam^0\pS$. Such a chirality violating
term has been introduced by Wilson \cite{Wilson:1975id}
to raise the masses of the unwanted doublers to values
of order of the cutoff, thereby decoupling them from
continuum physics.

As discretized supercharge \refs{wess32} we take
\eqnl{
Q^{(1)}=(\pi,\psi)+\ui(\phi,\hF^0\psi)+\ui(W',\gam^0\psi).}{lmod14}
A careful calculation yields the following  anticommutation
relations,
\eqnl{
\h\{ Q_\al^{(1)} , Q_\beta^{(1)} \} =
(\slashed{P} \gam^0)_{\al \beta}
-\ui (\gam^1)_{\al\beta} (W',\pA\phi)-\delta_{\al\beta}(W',\pS\phi)}
{lmod15}
with energy and momentum
\eqngrl{
2P_0&=&(\pi,\pi)-(\phi,\triangle\phi)+
\left(W',W'\right) + \left(\psi,\hF \psi\right),}
{2P_1&=&2\big(\pA\phi,\pi\big)
-\left(\psi,\gam_*\hF^0\psi\right),\quad
\hF=\hF^0+\gam^0 W''.}{lmod16}
To arrive at these results one uses the identity
\eqnn{
(\pi,\p\phi)+\ui (\p^\dagger\psi_1,\psi_1)
=(\p\phi,\pi)-\ui(\psi_1,\p^\dagger\psi_1),}
which holds for any real difference operator $\p$.
The superalgebra can be rewritten as
\eqnl{
\h\{ Q^{(1)} , \bar Q^{(1)} \} =\slashed{P}
+\ui \gam_*(W',\pA \phi)-\gam^0 (W',\pS\phi).}{lmod18}
The last term is absent in the superalgebra
\refs{algebra2} and breaks Lorentz covariance
explicitly. This lattice artifact originates in
the Wilson term $-\gam^0\pS$ in \refs{lmod13a}.
This term must vanish in the continuum
limit. One may wonder whether
there exist other improvement terms we could add to
a local $-\ui\gam_*\pA$ in order to
avoid the fermion doubling. However, since
for Majorana fermions the terms
\eqnn{
(\psi,\pS\psi),\quad
(\psi,\gam^1\pS\psi),\quad (\psi,\gam_*\pS\psi)}
are constant or zero,
all terms but $\gam^0\pS$ do not show up
in the right hand side of \refs{lmod16}
and we obtain the same result as if we
had chosen $\hF=-\ui\gam_*\pA$.
Hence only the Wilson term $\sim\gam^0\pS$ 
can be used to avoid the fermion doubling.
This argument does not apply to theories
with several Majorana fermions and in particular
to models with extended supersymmetry. 

Models with $\mN= 2$ supersymmetry contain the
second supercharge in \refs{wess40}, the lattice
version of which reads
\eqnl{
Q^{(2)}=(\pi,I\psi)+\ui(\phi,\hF^0 I\psi)-\ui(W',\gam^0 I\psi),}{lmod19}
and satisfies the same anticommutation relations
as $Q^{(1)}$, up to a sign change of the last two
terms in \refs{lmod18}. The anticommutator of
two lattice charges reads
\eqnl{
\h\{ Q^{(I)} , \bar Q^{(J)} \}=
\delta^{IJ}\slashed{P}+\ui\gam_* \mZ_S^{IJ}+\mZ_L^{IJ},}{lmod21}
where the `would-be' central charges
\eqnl{
\mZ_S^{IJ}=\sigma_3^{IJ}(W',\pA \phi)-(\sigma_1)^{IJ}(U',\pA\phi)}{lmod23}
approach the central charges \refs{wess42} of
the continuum model.  To arrive at \refs{lmod21} one needs
the harmonicity of the superpotential which
in turn implies the existence of a function $U(\phi)$ with $IW'=U'$,
and this function enters the central charges.
However, since the \emph{Leibniz rule}
never holds on the lattice, the integrands
$W'\cdot\pA\phi$ and $U'\cdot\pA\phi$ in \refs{lmod23}
are not just total derivatives as in the continuum
and as a consequence the terms
$\mZ_S^{IJ}$ are not central to the algebra.
The annoying terms
\eqnl{
\mZ_L^{IJ}= -(\sigma_3)^{IJ}\gam^0(W',\pS\phi)+
(\sigma_1)^{IJ}\left(\gam^0(U',\pS\phi)
-\ui (\pi,I\pS\phi)-\ih (\psi,I\pS\psi)\right)}{lmod25}
in \refs{lmod21}
are pure lattice artifacts and
vanish for antisymmetric lattice
derivatives.
\par
\textbf{Free Wess-Zumino model ($\mN=1$):}
For simplicity we consider the free model
with scalars of equal mass. The superpotential
reads $W=\h m \phi_a\phi^a$ and with $W'=m\phi$
the `would-be' central charge vanishes,
\eqnl{
(W',\pA\phi)=m(\phi,\pA\phi)=0.}{fwz1}
As Hamiltonian we \emph{choose}
the square of the supercharges,
\eqngrl{
H&=&\h \{ Q_1, Q_1\}=\h\{ Q_2, Q_2\}
=P_0-m(\phi,\pS\phi),}
{2P_0&=& (\pi ,\pi)
+ \left(\phi ,(-\triangle+m^2)\phi\right)
+\left(\psi,\hF \psi\right),}{fwz3}
where the Dirac-Hamiltonian for the
non-interacting model is just
\eqnl{
\hF=-\ui\gam_*\pA+\gam^0(m-\pS)
\mtxt{with}
\hF^2=(-\triangle+m^2-2m\pS)\id_2}{fwz5}
and $-\triangle= \p\p^\dagger$.
For antisymmetric derivatives the pure lattice
artifacts containing $\pS$ vanish and
with $2P_1=\{Q_1,Q_2\}$ we obtain the familiar algebra 
\eqnl{
\{ Q_\al,Q_\beta \}= 2(\gam^\mu \gam^0 )_{\al \beta}
P_\mu,\quad [Q_\al,P_\mu]=0,\quad
[P_0,P_1]=0.}{fwz7}
We conclude that the $\mN= 1$
superalgebra in $1+1$ dimensions
can be represented as a \emph{free Wess-Zumino model}
on a space lattice.

\subsection{Lattice Derivatives}

At this point some words about lattice derivatives
are in order. At first instance one may
think that the local right- and left derivatives
\eqnl{
(\pR f)(n)=f(n+1)-f(n)\mtxt{and} (\pL f)(n)= f(n)-f(n-1)}
{lmod5}
are ideal candidates for a lattice derivative.
With respect to the $\ell_2$-scalar product of
two lattice functions,
\eqnl{
(f,g) =\sum_{n=1}^N f(n)g(n),}{lmod6}
the adjoint of the left-derivative is minus the
right-derivative, $\pL^\dagger=-\pR$.
Both derivatives share the property that
$(1,\pR f) = (1, \pL f)=0$. But the corresponding
momenta $\imL=-\ui\pL$ and $\imR=-\ui\pR$
are not Hermitian and possess complex eigenvalues,
\eqnn{
\lam_k(\imR)=\bar\lam_k(\imL)=2\ue^{\ui p_k/2}\sin \frac{p_k}{2},
\quad\mtxt{with}
p_k=2\pi k/N,
\quad\mtxt{and} k=1,\dots,N.}
If we insist on a Hermitian momentum we could
choose the \emph{antisymmetric} derivative operator
\eqnl{
\pRL=\h(\pR+\pL)=-\pRL^T}{lmod8}
which is used in many lattice calculations.
The $N$ real eigenvalues of $\imRL$ read
\eqnn{
\lam_k(\imRL)=\sin p_k=\Re\left(\lam_k(\imR)\right),}
and waves with the shortest wavelength, that
is with $p_k$ at the boundary of the first Brillouin
zone, are zero modes of $\pRL$.
Hence, by trying to preserve the hermiticity of $\hat p$ in
this naive way immediately introduces spurious
zero modes that are responsible for the fermion
doubling problem.\\
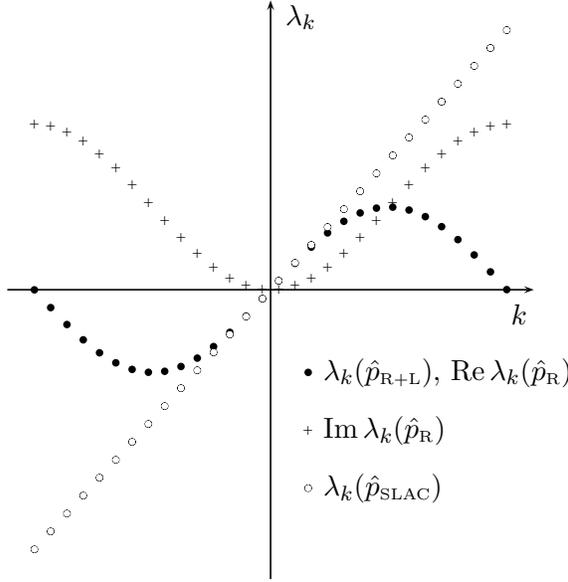
\begin{minipage}[t]{7.6cm}
\psset{xunit=10mm,yunit=11mm,linewidth=0.6pt,dotsize=1mm,plotpoints=30}
\begin{pspicture}(-3.6,-3.6)(3.5,3.5)
\psline{->}(-3.5,0)(3.5,0)
\rput(3.3,-0.3){$k$}
\psline{->}(0,-3.5)(0,3.5)
\rput(0.4,3.3){$\lam_k$}
\psplot[plotstyle=dots,showpoints=true]{-3.14}{3.14}
{57.3 x mul sin}
\psplot[plotstyle=dots,dotstyle=+,showpoints=true]{-3.14}{3.14}
{28.65 x mul sin 2 exp 2 mul}
\psplot[plotstyle=dots,dotstyle=o,showpoints=true]{-3.14}{3.14}{x}
\psdot(.5,-1)
\psdot[dotstyle=+](.5,-1.7)
\psdot[dotstyle=o](.5,-2.4)
\rput[l](.7,-1){$\lam_k(\imRL),\,\Re\lam_k(\imR)$}
\rput[l](.7,-1.7){$\Im\lam_k(\imR)$}
\rput[l](.7,-2.4){$\lam_k(\imSL)$}
\end{pspicture}
\end{minipage}
\begin{minipage}[b]{7cm}
A third alternative for the lattice momentum is the Hermitian
and nonlocal SLAC operator $\imSL= -\ui\pSL$
introduced by Drell, Weinstein and Yankielowicz
\cite{Drell:1976bq} with
real eigenvalues $p_k$.
This operator has no spurious mirror
states in the first Brillouin zone. The
eigenvalues of $\imSL$
are eigenvalues of the momentum
operator in the continuum and the difference
between the lattice and continuum results
are minimized. In the figure on the left
we have plotted the eigenvalues of the
various lattice operators. The real parts
of the eigenvalues of $\imR$ and $\imL$
are just the eigenvalues of $\imRL$. The
eigenvalues of $\imRL$ are twofold degenerate. The
SLAC operator has the same dispersion relation
as the momentum in the continuum.
\end{minipage}\par
Besides $\pR,\pL,\pRL$ and $\pSL$ there are
many other local and nonlocal candidates for
lattice derivatives with the correct
naive continuum limit. However, it is easy to see
that no linear difference operator
will obey the Leibniz rule. Many problems
in supersymmetric lattice theories
are exactly due to this fact, see \cite{Dondi:1977tx}.

In order to better understand the dependency of
the spectrum and doubling phenomenon on the lattice
derivative we consider the following one-parameter interpolating
family of \emph{ultra-local} difference operators
\eqnl{
\p_\al=\h(1+\al)\pR+\h(1-\al)\pL=
\pS+\pA,}{fwz9}
with symmetric and antisymmetric parts
\eqnl{
\pS=\h\al(\pR-\pL)=\h\al\pR\pL\mtxt{and}
\pA=\h(\pR+\pL)=\pRL.}{fwz11}
When the parameter $\al$ varies from
$1$ to $-1$, then $\p_\al$ interpolates between
$\pR$ and $\pL$. For $\al=0$ we obtain the
antisymmetric operator $\pA$ in \refs{lmod5}. 

The $2N$ eigenvalues of the Hermitian \emph{Dirac-Hamiltonian}
\refs{fwz5} depend on the deformation parameter as follows,
\eqnl{
\lam_k(\al)=\lam_{N-k}(\al)=\pm \sqrt{m^2+4\al(\al\ps m)\sin^2(\h p_k)
+(1\ms\al^2)\sin^2(p_k)},}{fwz15}
where $p_k=2\pi k/N$ and $k$ runs from $0$ to $N\ms 1$.
For the extreme
cases $\al= 0,1$ we obtain\par \vskip-.1cm
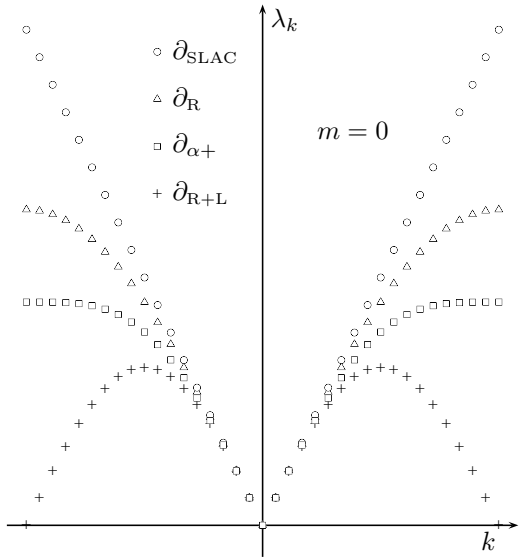
\begin{minipage}[t]{7.6cm}
\psset{xunit=10mm,yunit=21mm,linewidth=0.6pt,dotsize=1mm,plotpoints=37}
\begin{pspicture}(-3.4,-.3)(3.4,3.5)
\psline{->}(-3.4,0)(3.4,0)
\psline{->}(0,-.2)(0,3.3)
\psplot[plotstyle=dots,dotstyle=+]{-3.14}{3.14}
{57.296 x mul sin 2 exp sqrt}
\psplot[plotstyle=dots,dotstyle=triangle]{-3.14}{3.14}
{28.648 x mul sin 2 exp 4 mul sqrt}
\psplot[plotstyle=dots,dotstyle=o]{-3.14}{3.14}
{x abs}
\psplot[plotstyle=dots,dotstyle=square]{-3.14}{3.14}
{57.296 x mul sin 2 exp 0.5 mul 28.648 x mul sin 2 exp 2 mul add sqrt}
\psdot[dotstyle=o](-1.4,3)
\psdot[dotstyle=triangle](-1.4,2.7)
\psdot[dotstyle=square](-1.4,2.4)
\psdot[dotstyle=+](-1.4,2.1)
\small
\rput(0.3,3.2){$\lam_k$}
\rput[l](-1.2,3){$\pSL$}
\rput[l](-1.2,2.7){$\pR$}
\rput[l](-1.2,2.4){$\p_{\al+}$}
\rput[l](-1.2,2.1){$\pRL$}
\rput(1.2,2.5){$m=0$}
\rput(3,-.1){$k$}
\end{pspicture}
\end{minipage}
\begin{minipage}[b]{7cm}
the eigenvalues
\eqnn{
\lam_k(0)=\pm \sqrt{m^2+\sin^2(p_k)}}
with multiplicity $4$ and
\eqnn{
\lam_k(1)=\pm\sqrt{m^2+4(1+ m)\sin^2(\h p_k)}}
with multiplicity $2$.
This should be compared with the eigenvalues
on the \emph{continuous} interval of 'length' $N$,
\eqnl{
\lam_k=\pm\sqrt{m^2+p_k^2}}{fwz21}
with multiplicity $2$.
One can show that for $\al$ greater then $\al_+$
or less then $\al_-$, where
\eqnn{
4\al_\pm =\pm \left(\sqrt{m^2+8}\mp m\right),}
\end{minipage}\par
\emph{all} eigenvalues have the same multiplicity
as in the continuum.
In particular, for massless
fermions there are no doublers for $\al^2>1/2$.
However, for $\al\in [\al_-,\al_+]$ some eigenvalues
have multiplicity four.
In the above figure we have plotted the
\emph{positive} eigenvalues of $\hF$ for $\al=0,1,\al_+$.
For comparison we have depicted the
positive eigenvalues of $\hF$ for the nonlocal SLAC derivative
\eqnl{
(\pSL)_{n\neq n'}=(-)^{n-n'}\frac{\pi/N}{\sin\big(\pi(n-n')/N\big)}\mtxt{and}
(\pSL)_{nn}=0.}{fwz23}
Despite being nonlocal the SLAC derivative has many
advantages as compared to the local operators
$\pR,\pL$ or $\pRL$: 
it is \emph{antisymmetric} such that for massless
fermions chiral symmetry is preserved.
By construction the $2N$ real eigenvalues of
$\hF=-i\gam_*\pSL+\gam^0m$ are identical to the $2N$ lowest
eigenvalues of the continuum operator on the
interval of `length' $N$, \refs{fwz21}.
For this reason $\pSL$ has been called
\emph{ideal lattice operator} in the literature. 
We do not expect that unwanted nonlocal
counterterms \cite{Karsten:1979wh} are required for the
two-dimensional supersymmetric Wess-Zumino
models. This is certainly the case for the
finite models with extended supersymmetry.
For the model with $\mN= 1$ supersymmetry
the same should be true since it does not
contain gauge fields which couple to high
momentum modes at the edge of the Brillouin
zone. Indeed, in \cite{Campos:2002zz} is has
been claimed that $\pSL$
approaches an ultra-local operator when
$N$ tends to infinity, except for a border
matrix. In the appendix we give
a detailed analysis of this interesting operator.

\subsection{On the Quality of Lattice Derivatives in Supersymmetric QM}

It is enlightening to retreat to quantum-mechanical
systems and study the supercharges
\eqnl{
Q=\begin{pmatrix}
0&A\cr A^\dagger&0\end{pmatrix},
\mtxt{with}
A=\p+W,\quad A^\dagger=\p^\dagger+W,}{sqm1}
and in particular the quality of lattice
approximations for different lattice derivatives
$\p$ in $A$. The supercharge squares to
\eqnl{
Q^2=\begin{pmatrix}AA^\dagger&0\cr 0&A^\dagger A\end{pmatrix},}{sqm3}
with isospectral discretized Schr\"odinger operators
\eqngrl{
AA^\dagger&=&\p\p^\dagger+\p W+W\p^\dagger+W^2}
{A^\dagger A&=&\p^\dagger\p+\p^\dagger W+W\p+W^2.}{sqm5}
They have identical spectra, up to possible
zero modes. If the Leibniz rule held on the
lattice, if $\p$ was antisymmetric and if we
could replace $\p W$ by $W'+W\p$, then we would find the
super-Hamiltonian of supersymmetric quantum
mechanics in the continuum,
\eqnl{
H=\begin{pmatrix}\p \p^\dagger+W'+W^2&0\cr 0&\p^\dagger\p -W'+W^2\end{pmatrix}
.}{sqm7}
The difference between $Q^2$ and $H$ is the analog
of the last two terms in \refs{lmod18} and
the difference in their spectra is a good measure
for the suitability of the chosen lattice derivative
as regards supersymmetry and
the speed with which the continuum limit is approached.
In the following figure we have plotted the eigenvalues
of $Q^2$ and $H$ for $\p=\pSL$, denoted by $Q^2_{\rm SLAC}$
and $H_{\rm SLAC}$ and for $\p=\pR$, denoted by
$Q^2_{\rm naiv}$ and $H_{\rm naiv}$. We took the
superpotential $W=\lam x^2$ which gives rise to
the supersymmetric anharmonic oscillator. 
\par
\begin{figure}[ht]
\psset{xunit=1mm,yunit=50mm}
\hskip5mm
\begin{pspicture}(-5,0.5)(130,1.9)
\psline[linestyle=dotted]{->}(0,1)(124,1)
\rput(122,0.93){$n$}
\psline{->}(0,.5)(0,1.8)
\rput(10,1.75){$E_n/E_n^{\rm exact}$}
\multirput(20,0.98)(20,0){6}{\psline(0,0)(0,0.04)}
\rput(50,1.5){$V_\pm=\lam^2 x^4\pm 2\lam x$}
\pscurve[showpoints=true,linecolor=blue]
(1,1)
(3,1)(5,1)(7,1)(9,1)(11,1)(13,1)
(15,1)(17,1)(19,1)(21,1)(23,1)(25,1)
(27,1)(29,1)(31,1)(33,1)(35,1)(37,1)
(39,1)(41,1)(43,1)(45,1)(47,1)(49,1)
(51,1)(53,1)(55,1)(57,0.977)(59,0.955)(61,0.957)
(63,0.958)(65,0.959)(67,0.960)(69,0.962)(71,0.963)(73,0.964)
(75,0.965)(77,0.966)(79,0.966)(81,0.967)(83,0.968)(85,0.970)
(87,0.973)(89,0.984)(91,0.987)(93,1.015)(95,1.016)(97,1.059)
(99,1.058)(101,1.115)(103,1.110)(105,1.180)(107,1.173)(109,1.254)
(111,1.244)(113,1.335)(115,1.322)
\pscurve[showpoints=true,linecolor=green]
(1,1)
(3,1)(5,1)(7,1)(9,1)(11,1)(13,1)
(15,1)(17,1)(19,1)(21,1)(23,1)(25,1)
(27,1)(29,1)(31,1)(33,1)(35,1)(37,1)
(39,1)(41,1)(43,1)(45,1)(47,1)(49,1)
(51,1)(53,1)(55,1)(57,1)(59,1)(61,1)
(63,1)(65,1)(67,1)(69,1)(71,1)(73,1)
(75,1)(77,1)(79,1)(81,1)(83,1)(85,0.999)
(87,1.001)(89,0.997)(91,0.999)(93,1.014)(95,1.023)(97,1.054)
(99,1.065)(101,1.107)(103,1.118)(105,1.171)(107,1.181)(109,1.243)
(111,1.253)(113,1.324)(115,1.332)
\pscurve[showpoints=true,linecolor=red]
(1,1.032)
(3,1.048)(5,1.066)(7,1.078)(9,1.087)(11,1.094)(13,1.099)
(15,1.101)(17,1.103)(19,1.103)(21,1.102)(23,1.099)(25,1.095)
(27,1.091)(29,1.085)(31,1.078)(33,1.069)(35,1.060)(37,1.049)
(39,1.037)(41,1.023)(43,1.007)(45,0.990)(47,0.962)(49,0.956)
(51,0.911)(53,0.936)(55,0.891)(57,0.933)(59,0.891)(61,0.943)
(63,0.904)(65,0.965)(67,0.926)(69,0.995)(71,0.958)(73,1.034)
(75,0.997)(77,1.080)(79,1.043)(81,1.133)(83,1.096)(85,1.192)
(87,1.156)(89,1.258)(91,1.221)(93,1.330)(95,1.293)(97,1.409)
(99,1.371)(101,1.493)(103,1.455)(105,1.584)(107,1.545)(109,1.681)
(111,1.641)(113,1.784)(115,1.743)
\pscurve[showpoints=true]
(1,0.991)
(3,0.993)(5,0.989)(7,0.983)(9,0.977)(11,0.971)(13,0.964)
(15,0.956)(17,0.948)(19,0.940)(21,0.932)(23,0.923)(25,0.913)
(27,0.904)(29,0.894)(31,0.884)(33,0.873)(35,0.862)(37,0.851)
(39,0.839)(41,0.828)(43,0.815)(45,0.803)(47,0.790)(49,0.776)
(51,0.762)(53,0.747)(55,0.732)(57,0.716)(59,0.699)(61,0.682)
(63,0.662)(65,0.640)(67,0.645)(69,0.625)(71,0.642)(73,0.634)
(75,0.654)(77,0.656)(79,0.676)(81,0.687)(83,0.708)(85,0.728)
(87,0.747)(89,0.776)(91,0.795)(93,0.831)(95,0.849)(97,0.894)
(99,0.911)(101,0.963)(103,0.979)(105,1.039)(107,1.053)(109,1.122)
(111,1.134)(113,1.211)(115,1.222)
\rput(80,0.81){$Q^2_{\rm SLAC}$}
\psline{->}(78,0.86)(79,0.96)
\rput(67,1.20){$H_{\rm SLAC}$}
\psline{->}(66,1.13)(67,1)
\rput(100,1.65){$Q^2_{\rm naiv}$}
\rput(85,0.63){$H_{\rm naiv}$}
\small
\rput(20,1.075){$20$}
\rput(40,1.075){$40$}
\rput(60,1.075){$60$}
\rput(80,1.075){$80$}
\rput(120,1.075){$120$}
\small
\psline(-2,0.8)(0,0.8)
\rput(-5,0.8){$0.8$}
\psline(-2,0.6)(0,0.6)
\rput(-5,0.6){$0.6$}
\psline(-2,1.2)(0,1.2)
\psline(-2,1)(0,1)
\rput(-5,1){$1.0$}
\rput(-5,1.2){$1.2$}
\psline(-2,1.4)(0,1.4)
\rput(-5,1.4){$1.4$}
\end{pspicture}
\caption{\label{susyop}\textsl{
Eigenvalues of $Q^2$ and $H$ for the
SLAC derivative and the right-derivative
for $N=180$ lattice points, length $L=30$
and $\lam=1$.}}
\end{figure}
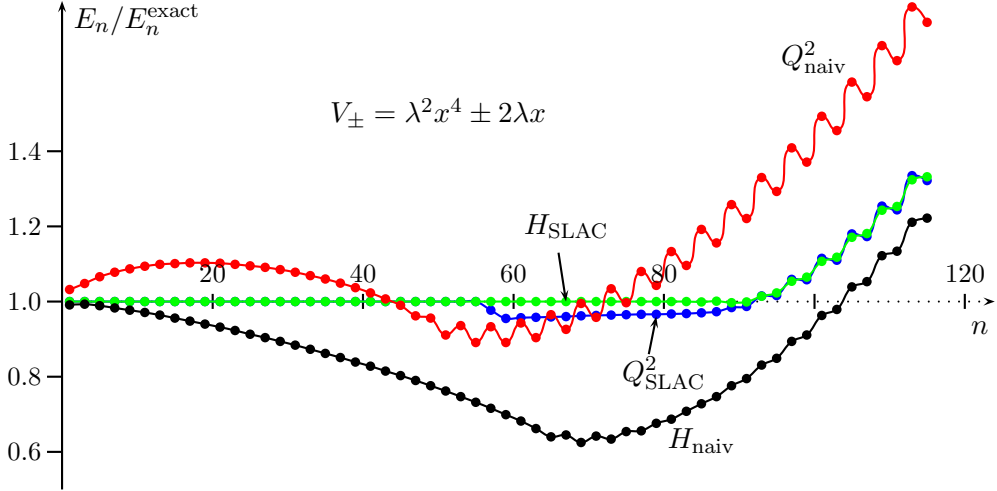
The lowest 57 eigenvalues of $Q^2$ and $H$ are
almost identical for the SLAC derivatives and the lowest
$90$ eigenvalues of $H_{\rm SLAC}$ agree
with the exact values (calculated on a much
finer grid). These results clearly demonstrate
the high precision of the SLAC derivative
in low-dimensional supersymmetric systems.
It does not matter whether we discretise
the supercharge or the super-Hamiltonian
as long as we choose the SLAC derivative.
After this detour to quantum mechanics we now
return to supersymmetric field theories.

\section{From the Dirac Operator to the Lattice $\mN=1$ WZ Model}

In this section we relate the
supercharges and Hamiltonians of two-dimensional
Wess-Zumino models on a spatial lattice to
suitable Dirac operators. We shall use the results
in \cite{Kirchberg:2004za} on the (extended) supersymmetries of
$\ui\di$ in arbitrary dimensions, specialized to
flat space and perform a dimensional
reduction such that the supercharges
of the lattice models can be related to
the reduced $\ui\di$. 

To see how Dirac operators relate to multi-dimensional
supersymmetric matrix-Schr\"odin\-ger operators
we generalize the reduction of the two-dimensional
Dirac equation to the Nicolai-Witten operator
to higher dimensions. For that purpose we dimensionally
reduce the Euclidean operator
\eqnl{
\ui \slashed{D}=\ui \Gamma^\mu D_\mu\,,\qquad
D_\mu=\p_\mu+ \ui A_\mu\,, \quad
{\Gamma^\mu}^\dagger = \Gamma^\mu,\quad \mu=1,\dots,2N,}{reda1}
from the products of cylinders,
\eqnl{
M=\underbrace{\R \times \ldots \times \R}_{\text{$N$-times}}
\times \underbrace{S^1 \times \ldots \times S^1}_{\text{$N$-times}}
=\R^N\times T^N}{reda3}
to the factor $\R^N$. 
We take $x^1,\dots,x^N$ as coordinates on $\R^N$
and $\theta^1,\dots,\theta^N$ as coordinates on the 
torus $T^N$, respectively. We
dimensionally reduce by assuming that the
Abelian gauge potential is independent of
the angles $\theta^n$. Then the Dirac operator
commutes with the (angular)momenta $-\ui\p_{\theta^n}$
and we may set $\p_{\theta^n}=0$.

\subsection{Reduction to Models with $\mN=1$}
If we further set $A_1(x)= \dots =A_N(x)=0$ and
define $\bar n= N\ps n$, then the square of the reduced Dirac
operator takes the form
\eqnl{
-\slashed{D}^2 \equiv 2H= p_n p_n + A_{\bar n}
A_{\bar n}-\ui \Gamma^n \Gamma^{\bar m} \p_n A_{\bar m},\mtxt{where}
p_n=\frac{1}{\ui}\frac{\p}{\p x^n}.}{reda5}
Note that the reduced operator $H$ contains no
first-derivative terms. It can be identified
with the Hamiltonian of a two-dimensional $\mN=1$
WZ-model on a spatial lattice, if we set
\eqnl{
x_n=\phi(n),\quad p_n=\pi(n)\mtxt{and}
\begin{pmatrix} \Gamma^n \cr\Gamma^{\bar n}\end{pmatrix}=\sqrt{2}\psi(n)
.}{reda7}
It follows with \refs{wess26} that
$(\psi,\gam^0\psi)=-\ui \Gamma^n \Gamma^{\bar n}$
holds true. If we further assume that the non-vanishing
components of $A_\mu$ have the form
\eqnl{
A_{\bar n}=-\big(\p^\dagger\phi\big)(n) + W'(\phi(n))\mtxt{with}
\p^\dagger=\pS-\pA,}{reda9}
then we find
\eqnl{
-\frac{1}{\sqrt{2}}\,\ui\di=
\underbrace{(\pi,\psi_1)+(W',\psi_2)}_{A_0}
\underbrace{-(\phi,\p\psi_2)}_{A_1}
=Q^{(1)}_1}{reda10}
with $Q^{(1)}$ in \refs{lmod14}. We conclude that
the Hamiltonian reads
\eqnl{
H= P_0+(W',\pA\phi)-(W',\pS\phi),}{reda11}
with $P_0$ from \refs{lmod16}.
Thus we have proved that the
super-Hamiltonian of the $\mN= 1$
Wess-Zumino model on a lattice with
$N$ sites is just the square of the Dirac
operator in $2N$ dimensions, dimensionally
reduced from $\R^N\times T^N$ to $\R^N$.

\subsection{Ground State of the Free Model}
For the massive non-interacting model we
have $2W= m\phi^2$. The corresponding
Hamiltonian is the sum of two commuting
operators, of the bosonic part
\eqnl{
\HB=\h(\pi,\pi)+\h (\phi,A^2\phi),\qquad
A^2=-\triangle +m\pS+m^2,}{reda13}
and the fermionic one
\eqnl{
\HF=\h(\psi,\hF\psi),\quad
\hF=-\ui\gam_*\pA+\gam^0(m-\pS).}{reda15}
We assume that the parameters are such
that $A^2$ is positive.
Near the continuum limit this is always
the case if the physical mass is positive.
The ground state wave function(al) of
the supersymmetric Hamiltonian factorizes,
\eqnn{
\Psi_0=\PsiB \PsiF\mtxt{with}
\HB\PsiB=\EB \PsiB\mtxt{and}
\HF\PsiF=\EF \PsiF.}
We choose the field representation for the scalar
field, such that
\eqnl{
\pi(n)=\frac{1}{\ui}\frac{\p}{\p\phi(n)}\mtxt{and}
\PsiB=\PsiB(\phi).}{reda17}
The bosonic factor $\PsiB$ is Gaussian
\eqnl{
\PsiB=\hbox{c}\cdot \exp\left(-\h (\phi,A\phi)\right)
\mtxt{and}\EB=\h\tr A.}{reda18}
Here $A$ is the positive root of the positive
and Hermitian $A^2$ in \refs{reda13}.
For the family of operators in \refs{fwz9}
the trace of $A$ is just half the sum
of the positive eigenvalues in \refs{fwz15}.

To find $\PsiF$ we introduce the (two-component)
eigenfunctions $v_k$ of $\hF$ with positive eigenvalues.
Since the Hermitian matrix $\hF$ is imaginary
the $v_k$ cannot be real and we have
\eqnl{
\hF v_k=\lam_k v_k\Longleftrightarrow
\hF \bar v_k=-\lam_k \bar v_k\quad (\lam_k>0).}{reda22}
The eigenvectors are orthogonal
with respect to the \emph{Hermitian} scalar product,
\eqnl{
(v_k,v_{k'})=\sum_{n,\al=1,2} \bar v_{k\al}(n)v_{k'\al}(n)=\delta_{kk'}
\mtxt{and}
(\bar v_k,v_{k'})=0.}{reda24}
Now we expand the Majorana spinors in terms of
this orthonormal basis,
\eqnl{
\psi(n)=\sum_{k=1}^N
\left(\chi_k v_k(n)+\chi^\dagger_k \bar v_k(n)\right)
,\mtxt{where}\chi_k=(v_k,\psi),\;\;
\chi^\dagger_k=(\bar v_k,\psi)}{reda26}
are one-component complex objects with
anticommutation relations
\eqnl{
\{\chi_k,\chi_{k'}\}=0\mtxt{and}
\{\chi_k,\chi^\dagger_{k'}\}=\delta_{kk'}.}{reda28}
Inserting the expansion \refs{reda26} into $\HF$ yields
\eqnl{
\HF=\ft12\sum_{k:\lam_k>0} \lam_k \left(\chi^\dagger_k\chi_k
-\chi_k\chi^\dagger_k\right).}{reda30}
It follows that the ground state of $\HF$ is
the Fock vacuum which is annihilated by all
annihilation operators $\chi_1,\dots,\chi_N$ and has energy
\eqnl{
\EF=-\ft12\sum_{k:\lam_k>0}\lam_k.}{reda32} 
Since $\hF^2=\id_2\otimes A^2$ we conclude,
that the positive eigenvalues of $\hF$ are
identical to the eigenvalues of $A$ such that
\eqnn{
E=\EB+\EF=0.}
Since $\Psi_0$ is normalizable for $A>0$
we see that the Hamiltonian admits a supersymmetric ground state for
\emph{all choices of the lattice derivative} $\p$,
provided $A$ is positive.

\subsection{Ground State for Strong Coupling}

In the extreme case of very strong self-coupling
of the scalar field we may neglect the derivative
term in the supercharge \refs{reda10} \cite{Elitzur:1982vh}.
Then $Q$ and $H$ are the sum of $N$ identical and
commuting quantum mechanical operators,
each defined on a given lattice site. Hence,
the ground state is a product state,
$\Psi_0(\phi)=\otimes_n \psi_0(\phi_n)$.
The operators on a \emph{fixed lattice site} read
\eqnl{
Q_1^{(1)}=\frac{1}{\ui}\psi_1\frac{\p}{\p\phi}+\psi_2W'(\phi)
\mtxt{and}
H=-\frac{\p^2}{\p\phi^2}+W'^2-\ui \psi_1\psi_2\,W''.}{reda40}
A normalizable zero-energy state is annihilated
by $Q_1^{(1)}$,
\eqnl{
\psi_0(\phi)=\ue^{-\ui\psi_1\psi_2 W(\phi)}\omega_0,}{reda42}
where $\omega_0$ is a constant two-component spinor. It is
well-known \cite{Jaffe:1987nx} and follows at once from
\refs{reda42} that supersymmetry is
unbroken if $p$ in
\eqnl{
W=\frac{1}{p}\lam\phi^p+ O(\phi^{p-1}),\quad \lam\neq 0,}{reda44}
is even and it is broken if $p$ is odd.
Note that $-\ui\psi_1\psi_2$
is Hermitian and has eigenvalues $\pm 1$ and that for even $p$
the state $\psi_0$ is normalisable if
\eqnn
{\ui\,\sign(\lam) \psi_1\psi_2\omega_0=\omega_0\mtxt{or}
\left(\psi_1-\ui\;\sign(\lam)\psi_2\right)\omega_0=0.}
To summarize, for even $p$ the $\mN= 1$ Wess-Zumino
model on the spatial lattice has always exactly
one normalizable zero mode in the strong-coupling limit.
For $\lam>0$ this product state has the form
\eqnl{
\Psi_0(\phi)=\exp\left(-\sum_{n=1}^N W(\phi_n)\right)
\Omega_0,\qquad
\big(\psi_1(n)-\ui\psi_2(n)\big)\Omega_0=0,\;\;\forall\; n.}{reda48}
In particular, for $\phi^4$-models supersymmetry
is broken in the strong-coupling limit where\-as it is
unbroken for $\phi^6$-models.

\section{From the Dirac Operator to the Lattice $\mN=2$
WZ Model}
It is known, that on flat spacetime the Euclidean
Dirac operator admits two supersymmetries if the
field strength commutes with an antisymmetric and
orthogonal matrix $I$, which defines a complex
structure \cite{Kirchberg:2004za}. The two real supercharges
\eqnl{
Q_1=\frac{1}{\sqrt{2}}\,\ui\,\Gamma^\mu D_\mu\mtxt{and}
Q_2=\frac{1}{\sqrt{2}}\,\ui\,I^\mu_{\;\;\nu}\Gamma^\nu D_\nu}{redb1}
form the superalgebra
\eqnl{
\h \{ Q_i , Q_j \}= \delta_{ij} H.}{redb3}
They can be combined to a \emph{nilpotent} complex charge
\eqnl{
Q=\frac{1}{\sqrt{2}}\left(Q_1+\ui Q_2\right)}{redb5}
and its adjoint $Q^\dagger$, in terms of which
the supersymmetry algebra takes the form
\eqnl{
H=\h\{Q,Q^\dagger\},\quad Q^2=Q^{\dagger 2}=0
\mtxt{and}[Q,H]=0.}{redb7}
To obtain $\mN= 2$ lattice models
on $N$ sites we consider the Dirac operator on
\eqnl{
M=\underbrace{\R \times \ldots \times \R}_{\text{$2N$-times}}
\times \underbrace{S^1 \times \ldots \times S^1}_{\text{$2N$-times}}
=\R^{2N}\times T^{2N}}{redb8}
in contrast to the $2N$-dimensional space in \refs{reda3}. 
Since the field strength
commutes with the complex structure $I$ it is very
convenient to introduce the corresponding
complex coordinates on $M$,
\eqnl{
z^n=x^n+\ui x^{\bar n}=x^n+\ui\theta^n,\quad
\bar n=2N+n,\quad n,\bar n\in\{1,\dots,2N\},}{redb9}
and fermionic annihilation and creation operators,
\eqnl{
\psi^n=\h\left(\Gamma^n+\ui\Gamma^{\bar n}\right),\quad
\psi^{\dagger n}=\h\left(\Gamma^n-\ui\Gamma^{\bar n}\right)\mtxt{with}
\{\psi^n,\psi^{\dagger m}\}=\delta^{mn}.}{redb11}
The condition that $F_{\mu\nu}$ commutes with
the complex structure $I$ implies the existence
of a real superpotential $\chi(z,\bar z)$, such that \cite{Kirchberg:2004za}
\eqnn{
Q= 2\ui \ue^{-\chi}\left(\sum_{n=1}^{2N}
\psi^n\frac{\p}{\p z^n}\right)\ue^\chi.}

\subsection{Reduction to Models with $\mN=2$}

Again we perform a dimensional reduction by assuming
that $\chi$ does not depend on the compact variables
$\theta^n$,
\eqnl{
\chi=\chi\left(x^1,\dots,x^{2N}\right)}{redb13}
and that the angular momenta $\p_{\theta^n}$ vanish.
In the sector with vanishing angular momenta the
complex charge simplify to
\eqnl{
Q= \ue^{-\chi} Q_0 \ue^\chi,\mtxt{where}
Q_0=\ui\sum_{n=1}^{2N}\psi^n\frac{\p}{\p x^n},}
{redb15}
since the complex $z^n$-derivative
becomes the real $x^n$-derivative in this
sector. This dimensional reduced supercharge
and its adjoint generate the superalgebra
\refs{redb7} with supersymmetric matrix-Schr\"{o}dinger 
operator 
\eqnl{
H=\h\{Q,Q^\dagger\}
=\underbrace{-\h\Delta +\h(\nabla \chi,\nabla\chi) + \h \Delta \chi}_{\HB}
-\underbrace{\sum \psi^{n\dagger} \chi_{,nm} \psi^m}_{\HF}.}
{redb17}
For example, for $\chi=-\lam r$ this is just the
Hamiltonian of the supersymmetrized Hydrogen atom
which has been introduced and solved in \cite{Kirchberg:2002me}.
It is evident from the representations \refs{redb15}
and \refs{redb17} that $Q$ decreases and
$Q^\dagger$ increases the eigenvalue of the number operator
\eqnl{
N=\sum \psi^{n\dagger}\psi^n}{redb19}
by one and $H$ commutes with $N$.
The eigenvalues of $N$ are $0,1,\dots,2N$.

As before, we interpret the $2N$ coordinates
$x^n$ and annihilation operators $\psi^n$
as values of two scalar and one Dirac field on
a one-dimensional lattice with $N$ lattice
sites. More precisely, we make the following identifications
for $n= 1,2,\dots,N$,
\eqngrl{
\phi(n)=\begin{pmatrix} x^{2n-1}\cr x^{2n} \end{pmatrix},&&
\pi(n)=\begin{pmatrix} p_{2n-1}\cr p_{2n} \end{pmatrix},}
{\psi(n)=\begin{pmatrix}\psi^{2n-1}\cr \psi^{2n} \end{pmatrix},&&
\psi^\dagger(n)=\begin{pmatrix}\psi^{\dagger\,2n-1}\cr \psi^{\dagger 2n}
 \end{pmatrix}.}
{redb25}
The free supercharge \refs{redb15} takes
the form
\eqnl{
Q_0=\ui\sum_{n=1}^N \psi(n)\frac{\p}{\p\phi(n)}\quad,\quad
Q_0^\dagger=\sum_{n=1}^N \psi^\dagger(n)\frac{\p}{\p\phi(n)}.}{redb26}
The remaining task is to find a superpotential
$\chi$ giving rise to interacting lattice Wess-Zumino
models. Since $\chi$ should be real we use
a representation for the two-dimensional
$\gam$-matrices such that $\ui\gam_*$ and $\gam^0$
are real,
\eqnl{
\gam^0=\sigma_3,\quad\gam^1=\ui\sigma_1,\quad\gam_*=-\sigma_2,}{redb27}
in order to obtain a real Dirac-Hamiltonian,
\eqnn{
\hF^m=-\ui \gam_*\p+m\gam^0=
\begin{pmatrix}m&\p\cr -\p&-m\end{pmatrix}.}
As explained above, $\p$ need not be anti-Hermitian
in which case we take
\eqnl{
\hF^m=\begin{pmatrix}m&\p\cr \p^\dagger&-m\end{pmatrix}
=-\ui\gam_*\pA+m\gam^0-\ui\gam^1\pS\mtxt{with}
(\hF^m)^2=(-\triangle+m^2)\id_2,}{redb29}
such that $\hF^m$ is real and Hermitian.
Note that the term containing $\pS$ is
not a momentum dependent mass term
as in \refs{lmod13a}. We have been lead to
a different type of Wilson term
as compared to the $\mN= 1$ model since
we have chosen a different representation
for the $\gam$-matrices. The earlier
Majorana representation \refs{wess26} is not
useful in the present context, since it would
lead to a complex $\chi$ in \refs{redb15}.

The term $\HF$ in \refs{redb17}
must contain the free Dirac-Hamiltonian
and this condition implies
\eqnl{
-\frac{\p^2\chi^m}{\p\phi_\al(n)\p\phi_\beta(n^\prime)}
=(\hF^m)_{\al\beta,nn^\prime},\quad \al,\beta=1,2, \quad
n,n^\prime=1,2,\dots,N.}{redb31}
Hence we expect that the real function
\eqnl{
\chi^m = -\h (\phi,\hF^m\phi),}{redb33}
is the superpotential for a $\mN= 2$
supersymmetric model. For these models we
use the following inner products
\eqnl{
(\phi,\tilde\phi)=\sum_{\al,n}
\phi_{\al,n}\tilde\phi_{\al,n}\mtxt{and}
(\psi,\tilde\psi)=\sum_{\al,n}
\psi_\al^\dagger(n)\tilde\psi_\al(n),
}{redb35}
for scalar doublets and Dirac spinors,
respectively. The corresponding supercharge
\eqnl{
Q=\ui \ue^{-\chi^m}Q_0\, \ue^{\chi^m}=\ui\sum_n \psi(n)\left(
\frac{\p}{\p\phi}-\hF^m\phi\right)}{redb36}
and its adjoint give rise to the following super Hamiltonian,
\eqnl{
\HB= -\h (\pi,\pi) + \h \big(\phi,(-\triangle+m^2)\phi\big),\quad
\HF= (\psi,\hF^m\psi).}{redb37}
Note that the superpotential $\chi^m$ is a harmonic
function, $\triangle\chi^m= 0$, and thus there
is no constant contribution to $\HB$.
The charges and Hamiltonian act on function(al)s
in the Hilbert space of the $\mN= 2$ lattice models
\eqnl{
\mH=\underbrace{\mh\otimes \cdots\otimes \mh}_{N-\text{times}}
,\mtxt{where} \mh=L_2(\R^2)\otimes \C^4}{redb39}
is the Hilbert space for the degrees of freedom
on one lattice site.

Now we turn to \emph{interacting models} 
by replacing the mass term in
\eqnn{
\chi^m=-\ft12 (\phi,\hF^0\phi)+\sum_n f\big(\phi(n)\big),\qquad
f(\phi)=\ft12 m(\phi_2^2-\phi_1^2),}
given by the quadratic harmonic function $f$,
by an arbitrary harmonic function $f(\phi)$ of the
two variables $\phi_1$ and $\phi_2$,
\eqnl{
\chi=-\ft12 (\phi,\hF^0\phi) + \sum f\big(\phi(n)\big), \mtxt{where}
\Delta f=0.}{redb41}
The supercharge and its adjoint are calculated
as
\eqnl{
Q=\ue^{-\chi}Q_0\ue^{\chi}\mtxt{and}
Q^\dagger =\ue^\chi Q_0^\dagger \ue^{-\chi}}{redb42}
with $Q_0$ and $Q_0^\dagger$ from \refs{redb26}.
After some algebra one finds for the bosonic
part of $H=\h\{Q,Q^\dagger\}$ the formula
\eqnl{
\HB =\h(\pi,\pi)-\h(\phi ,\triangle \phi)
+ \h\left(\frac{\p f}{\p\phi},\frac{\p f}{\p \phi}\right)
+\left(\frac{\p g}{\p \phi_1},\p^\dagger\phi_1\right)
-\left(\frac{\p g}{\p \phi_2},\p\phi_2\right)
}{redb43}
where the harmonic functions $f$ and $g$ are the real
and imaginary parts of an analytic function, such
that
\eqnl{
\frac{\p f}{\p\phi_1}=\frac{\p g}{\p\phi_2} \mtxt{and}
\frac{\p f}{\p\phi_2}=-\frac{\p g}{\p\phi_1}.}{redb48}
For the fermionic part of the Hamiltonian one obtains
\eqnl{
\HF
=(\psi,\hF^0\psi) - \left(\psi,\gam^0\Gamma(\phi)\psi\right),\qquad
\Gamma(\phi)=f_{,11}(\phi)
-\ui \gam_* f_{,12}(\phi).}{redb45}
The last term contains the Yukawa coupling
between scalar and Dirac fields.
Note that the last two terms in \refs{redb43} can be
rewritten as
\eqnl{
\mZ=-\left(\frac{\p g}{\p\phi},\pA\phi\right)+
\left(\frac{\p g}{\p\phi},\sigma_3\pS\phi\right).}{redb47}
In the continuum limit the first term on the right
becomes a surface term commuting with the supercharges
and the second term, which is a lattice artifact,
must vanish. Thus it is natural to set
\eqnl{
\HB+\HF= P_0+\mZ}{redb49}
and interpret the first term
\eqnl{
P_0=\h (\pi,\pi) - \h (\phi,\Delta \phi) + (\psi,h^0_\uF \psi)
+ \h \left( \frac{\p f}{\p\phi} , \frac{\p f}{\p\phi} \right)
- \left( \psi,\gam^0 \Gamma(\phi)\psi \right)}{redb51}
as energy and the second term as `would be' \emph{central charge}
$\mZ$ in \refs{redb47}. This agrees 
with our interpretation for solitonic configurations 
saturating the BPS-bound. To see that more clearly
we consider the energy of a purely bosonic static
solution,
\eqnl{
E=P_0=-\h(\phi,\Delta \phi)
+ \h \left( \frac{\p f}{\p\phi} , \frac{\p f}{\p\phi}\right).}{redb53}
From the very construction it is evident, that
there is a BPS-bound. One just adds the
non-negative operator $\HB$ in
\refs{redb43} to the non-negative operator one gets when
changing the signs of $f$ and $g$ and finds 
\eqnl{
E \geq |\mZ|.}{redb55}
For example, a cubic superpotential $f+\ui g=\lam\phi^3/3$
leads to a $\phi^4$-models with
\eqnl{
P_0=
\h (\pi,\pi)-\h (\phi, \Delta \phi)
+(\psi ,h^0_\uF \psi) + \ft12\lam^2\left( \phi,\phi \right)^2 -
\left( \psi,\gam^0\Gamma(\phi)\psi\right).}{redb59}
It contains a scalar and pseudoscalar Yukawa
interaction with
\eqnn{
\Gamma(\phi)=2\lam (\phi_1+\ui\gam_*\phi_2).}
The would-be central charge is cubic in the scalar
fields and reads
\eqnl{
\mZ=
2\lam \left(\phi_1\phi_2,\p^\dagger\phi_1 \right)
-\lam \left(\vphantom{\p^\dagger}\phi_1^2-\phi_2^2,\p\phi_2\right).}{redb63}
Before turning to the discussion of the ground
state we note, that the conserved number operator
\eqnl{
N=\sum_n \psi^\dagger(n)\psi(n)}{redb70}
leads to a decomposition of the Hilbert space
\refs{redb39} in orthogonal subspaces labelled
by the \emph{fermion number},
\eqnl{
\mH=\mH_0\oplus \mH_1\oplus\dots\oplus \mH_{2N-1}\oplus \mH_{2N},
\qquad N\big\vert_{\mH_p}=p\id.}{redb72}
The nilpotent supercharge $Q$ decreases $N$ by one,
$Q^\dagger$ increases it by one and the super-Hamiltonian
commutes with $N$,
\eqnl{
[N,Q]=-Q,\quad [N,Q^\dagger]=Q^\dagger\mtxt{and}
[N,H]=0.}{redb74}
We call the subspace $\mH_p$ $p$-particle sector.
The states in the zero-particle
sector are annihilated by $Q$ and those in the
$2N$-particle sector by $Q^\dagger$.

\subsection{Ground State of the Free Model}
The Hermitian lattice Dirac-Hamiltonian $\hF^m$ in \refs{redb29}
is real and can be diagonalized by an orthogonal matrix $S$,
\eqnl{
\hF^m = S^{-1} D S, \quad D=\text{diag}(d_1,d_2,\dots,d_{2N}).}{redc1}
We rotate the field-variables with $S$,
\eqnn{
\xi=S \phi, \quad \eta = S \psi \mtxt{and}
\eta^\dagger = S \psi^\dagger.}
The new fields still obey the standard
anticommutation relations, e.g.
\eqnl{
\{\eta^\dagger_\al(n),\eta_\beta(m) \}=
\delta_{\al\beta} \delta_{nm},}{redc3}
and the transformed supercharges read
\eqnl{
Q =\ui\eta\left(\frac{\p}{\p\xi} - D\xi\right)
\mtxt{and}
Q^\dagger=\ui\eta^\dagger \left(\frac{\p}{\p\xi}+ D\xi\right)}{redc5}
and show, that the new degrees of freedom decouple.
Hence the ground state must have the product form
\eqnl{
\Psi_0=\exp\left(-\h\sum |d_a|\xi_a^2\right) |\Omega \rangle}{redc7}
and the supercharges act on this state as follows,
\eqnl{
Q \Psi_0= 2\ui \sum_{a:\,d_a>0} d_a\eta_a\xi_a
\Psi_0\quad,\quad
Q^\dagger\Psi_0=-2\ui \sum_{a:\,d_a<0} d_a \eta^\dagger_a\xi_a \Psi_0.}{redc9}
This way we arrive at the following conditions for
this state to be invariant,
\eqnl{
d_a > 0 \Longrightarrow \eta_a |\Omega\rangle = 0 \mtxt{and}
d_a < 0 \Longrightarrow \eta^\dagger_a | \Omega \rangle = 0.}{redc11}
This leads to the \emph{unique} normalizable ground state \refs{redc7}
with
\eqnl{
\vert\Omega \rangle=\prod_{d_a<0}\eta^\dagger_a \va.}{redc13}
which is annihilated by the supercharges and hence
has vanishing energy. There are $N$ positive
and $N$ negative eigenvalues of $\hF^m$ such
that the invariant vacuum state lies in the
middle sector $\mH_N$ in the decomposition
\refs{redb72} of the Hilbert space. All fermionic
states with negative energies are filled. This
is just the Dirac-sea filling prescription.  
Note that our result is the lattice version
of the continuum result for the ground state,
\eqnn{
\Psi_0=\exp\left(-\h\int \phi_\al\sqrt{-\triangle+m^2}\phi_\al\right)
\,\vert\Omega \rangle.}

\subsection{Ground States for Strong Coupling}

In the strong-coupling limit we may neglect the
spatial derivatives such that the supercharges
and the Hamiltonian becomes the sum of $N$ commuting
operators, each defined on one lattice site
\cite{Elitzur:1983nj}. The operators \emph{on a
given site} take the form
\eqnl{
Q=\ui \psi\left(\nabla+\nabla f\right),\quad
H=-\h\triangle+
\h(\nabla f,\nabla f)
-\psi^\dagger f'' \psi.}{redc15}
Now we explicitly construct the ground state for the
harmonic superpotential
\eqnl{
f(\phi)=\frac{\lam}{p}\,\Re \phi^p,\qquad
\phi=\phi_1+\ui\phi_2=r \ue^{\ui\theta},}{redc17}
which gives rise to a supersymmetric \emph{anharmonic
oscillator} on the Hilbert space $\mh=L_2(\R^2)\times\C^4$.
The bosonic part of $H$ reads
\eqnl{
\HB=-\h\triangle +V\mtxt{with}
V=\h\lam^2 r^{2p-2},}{redc19}
and its fermionic part
\eqnl{
\HF=
\lam (p\ms 1)\,\psid\begin{pmatrix}-\Re \phi^{p-2}&\Im \phi^{p-2}\cr
\Im \phi^{p-2}&\Re\phi^{p-2}\end{pmatrix}\psi}{redc21}
It is useful to note that $H$ commutes with the operator
\eqnl{
J=L+S,\quad S=-s\psid \sigma_2\psi,\quad s=\h(p-2)}{redc23}
and that the ground state must reside in the
two-dimensional sector with particle number
$N=\psi^\dagger\psi= 1$, since the restriction
of $\HF$ to the zero- and two-particle sectors
vanish and $\HB>0$. The one-particle sector is
spanned by the following two eigenstates of $S$,
\eqnl{
\up=\left(\psid_1-\ui\psid_2\right)\vac\mtxt{and}
\down=\left(\psid_1+\ui\psid_2\right)\vac}{redc25}
with eigenvalues $1$ and $-1$, respectively. Here
$\vac$ denotes the Fock-vacuum which is annihilated
by the annihilation operators $\psi_\al$.
Diagonalising $J$ in this sector leads to the ansatz
\eqnl{
\psi_{0j}(\phi)=R_{j+}(r) e^{i(j-s)\theta}\up+R_{j-}(r)e^{i(j+s)\theta}
\down,}{redc27}
where the $J$-eigenvalue $j$ is integer for
even $p$ and half-integer for odd $p$.
Inserting into $Q\psi_j= Q^\dagger\psi_j= 0$
yields the following coupled system of
first order differential equations for
the radial functions
\eqnn{
R'_{j\pm}(r)-\frac{s\mp j}{r}R_{j\pm} (r)+\lam r^{p-1} R_{j\mp}(r)=0.}
The square integrable solutions are Bessel functions
\eqnl{
R_{j\pm} (r)=c\, r^{p-1}\,
\mbox{K}_{\ha\pm\frac{j}{p}}
\left(\textstyle\frac{\lam}{p}r^p\right)\mtxt{with}
j\in \{-s,-s+1,\dots,s-1,s\}.}{redc29}
The number of supersymmetric ground states of the
models with $\phi^{2p-2}$ self-interaction is 
just $p-1$. The $(p-1)^N$ normalizable invariant
eigenstates are
\eqnl{
\Psi_{0,j_1,\dots,j_N}=\bigotimes_{n=1}^N
\psi_{0j_n}(\phi_n)\in \mh_1\times\dots\times \mh_N.}{redc31}
For example, for the $\mN=2$ model with $\phi^4$
interaction there exist $2^N$ normalizable zero modes
in the strong-coupling limit. This number diverges
in the thermodynamic limit. On the other hand,
there is exactly one normalisable zero mode
when one switches off the self-interaction.
This discrepancy between the number of
supersymmetric ground states in the weak and
strong-coupling regimes becomes even more puzzling when
one takes into account certain rigorous theorems
on the stability of such states under analytic
perturbations discussed in the following section.
The zero modes in \refs{redc31} with
radial functions \refs{redc29} have been constructed
previously in \cite{Elitzur:1983nj,Jaffe:1987nx}
and \cite{Bruckmann:1997}.

\section{From Strong to Weak Couplings}

\subsection{Perturbation Theory and Zero Modes}
\label{breaking}
Let us recall a well known result for perturbation theory of zero modes
in supersymmetric quantum mechanics \cite{Jaffe:1987nx}.
We consider the $\mN=1$ case and denote the single Hermitian
supercharge by $Q_0$,
\eqnl{
Q_0^2=H_0, \quad \{\Gamma,Q_0\}=0,\quad \Gamma^\dagger=\Gamma,\quad
\Gamma^2=\id.}{hilfs1}
In addition, we define the projection operators
\eqnl{
\mP_\pm = \h (\id \pm \Gamma)}{project}
which project on the $\pm 1$ eigenspaces of $\Gamma$.
These eigenspaces are denoted by $\mH_{\uB/\uF}$.
In the following we assume that there are no zero
modes in $\mH_\uF$ and at least one zero mode  $\Psi_0$
in the bosonic sector $\mH_\uB$. We
perturb the operator $Q_0$ by an operator $\epsilon Q_1$ with real
parameter $\epsilon$, $Q(\epsilon) = Q_0 + \epsilon Q_1$, where
$\{ Q_1 , \Gamma \} =0$. We want to solve the
eigenvalue equation
\eqnn{
Q(\epsilon) \Psi(\epsilon) = \lambda(\epsilon) \Psi(\epsilon),}
with $\lambda(0) = 0$ and $\Psi(0) = \Psi_0$.
We consider the following formal power series in $\epsilon$,
\eqnn{
\Psi(\epsilon)= \Psi_0 + \sum_{k=1}^\infty \epsilon^k \Psi_k, \quad
\lambda(\epsilon)= \sum_{k=1}^\infty \epsilon^k \lambda_k.}
\textbf{Proposition:} Under the assumptions above one has
$\lambda(\epsilon) = 0$ and
$\Gamma \Psi(\epsilon) = \Psi(\epsilon)$ in the sense of formal power
series.

\emph{Proof by induction:} To order $\epsilon^0$ the proposition
holds. Let us assume it holds up to order $\epsilon^{j-1}$.
To order $\epsilon^j$ we obtain the equation
\eqnl{
Q_0 \Psi_j + Q_1 \Psi_{j-1} = \lambda_j \Psi_0.}{perturb3}
Taking the scalar product with $\Psi_0$ yields
\eqnn{
\lambda_j = (\Psi_0 , Q_1 \Psi_{j-1}).}
Since $\Gamma$ squares to $\id$ and anticommutes
with the perturbation we find
\eqnn{
\lambda_j = ( \Gamma^2 \Psi_0 , Q_1 \Psi_{j-1} ) = -
( \Gamma \Psi_0 , Q_1 \Gamma \Psi_{j-1} ) = - ( \Psi_0
, Q_1 \Psi_{j-1} ) = - \lambda_j,}
which proves that $\lambda_j=0$. Furthermore, with
\eqnn{
Q_0 \Gamma \Psi_j = - \Gamma Q_0 \Psi_j = \Gamma Q_1 \Psi_{j-1} = -
Q_1 \Psi_{j-1} = Q_0 \Psi_j,}
we conclude
\eqnn{
Q_0 \mP_-\Psi_j =0,}
where we used the projection operator $\mP_-$
introduced in \refs{project}.
As $\mP_- \Psi_j$ is a zero mode of $Q_0$ it
follows by assumption that it resides in $\mH_\uB$. But as
$\mP_-$ projects onto $\mH_\uF$, we conclude
$\mP_- \Psi_j=0$ or $\Psi_j \in \mH_\uB$. This
then proves our statement.
Note that the statement has been proved in the
sense of formal power series only. In case $\lambda(\epsilon)$
is not analytic at $\epsilon=0$ the result above
maybe misleading.

\subsection{The $\mN=1$ Case}
\label{n1weak}
In what follows, we compare the strong-coupling results with
the usual perturbation theory around minima of the potential.

In the case $\text{deg}(W)=p$ even, supersymmetry is never broken,
neither in the strong-coupling limit
nor in perturbation theory. For even $p$ there
is at least one minimum of the potential $V=\h(W')^2$ with
$V=0$. The quadratic approximation of the potential at the critical
points yields for each minimum one normalizable zero mode similar to
the ground state of the free model. In contrast to the strong-coupling
limit there may be more than one perturbative zero mode, but
they always come in an odd number. The difference of bosonic and
fermionic zero modes is $\pm1$ as in the strong-coupling limit.

In the case $\text{deg}(W)=p$ odd, the difference between the
strong-coupling limit and perturbation theory is more severe.
Supersymmetry is broken in the strong-coupling limit but it may be
unbroken in perturbation
theory. Let us consider an explicit example,
\eqnl{
W(\phi) = \frac{g_2}{2} \phi^3 + g_0 \phi.}{n1weak1}
Perturbation theory for $g_0<0$ predicts one bosonic and
one fermionic zero mode (unbroken supersymmetry), and
broken supersymmetry for $g_0 > 0$. The strong-coupling limit
states that supersymmetry is broken for all $g_0$.

In Appendix \ref{n1inequ} we provide the
rigorous proof that $\lambda A_1$
with $A_1$ given in \refs{reda10} is an \emph{analytic
perturbation} of $A_0$ in \refs{reda10}.
This implies that all eigenvalues are analytic functions of the parameter
$\lambda$. Assume now that in a finite range of the parameter $\lambda$
there is a ground state with energy zero. As an analytic
function which vanishes in some finite range is identically zero, the
number of zero modes changes at most at isolated points of the parameter
space of $\lambda$. Furthermore, in the strong-coupling limit, we have
either bosonic or fermionic zero modes. In subsection \ref{breaking} we
have proved
that under this assumption a zero mode always remains a zero mode.
We conclude that, generically, the number of zero modes is
given by the number of zero modes in the strong-coupling
limit. Generically, since for certain discrete values
of $\lambda$ the number of zero modes could be enhanced.
Moreover, as the index also depends analytically on the parameter
$\lambda$, we are able to calculate this index in the
strong-coupling limit.

In the continuum and infinite-volume limit these arguments may break
down, as the estimates necessary for proving analyticity
(see Appendix \ref{n1inequ}) may not be valid anymore.
In the \emph{unbroken case} we can definitely conclude
that the theory is still unbroken in the continuum and infinite-volume
limit. Suppose we know that for any finite lattice there is at least one
ground state with zero energy. As the limit of zero is again zero this
mode survives in the limit.
In the \emph{case of broken supersymmetry} a non-zero energy eigenstate may
become a zero mode in the continuum and infinite-volume limit, and
supersymmetry may get restored in this limit although it is
broken for all finite lattices.

Indeed, for negative $g_0$ in our example above, the scalar field has a
non-vanishing vacuum expectation value and 
therefore the fermionic field $\psi$ acquires a
non-zero mass. As there is no massless Goldstone
fermion, supersymmetry has to be unbroken in this case \cite{Witten:1981nf}.

Let us summarize. On a finite lattice, the
\emph{strong-coupling limit gives the
correct number of zero modes of the full problem.}
There is only one zero mode in the case
where $\text{deg}(W)=p$ is even, and otherwise there is
no zero mode. Variations of the parameters in the superpotential of
power less than $p$ do neither change the number of zero
modes nor the index. For example, in the model with superpotential given in
\refs{n1weak1}, it is impossible to have two phases of broken and
unbroken supersymmetry (depending on the value of the parameter $g_0$) on a
finite lattice. The numerical simulations in \cite{Beccaria:2004pa}
may be interpreted as hinting towards such a phase transition in
the continuum theory.

\subsection{The $\mN=2$ Case}
\label{n2weak}
Similar to the case $\mN=1$, we prove in Appendix \ref{n2inequ} that the
index in the strong-coupling limit is the same as for the full problem.
This implies
that we have at least $(p-1)^N$ zero
modes for the theory on finite lattices. For the continuum theory in a
finite volume, it was
shown using methods of constructive
field theory that the $\mN=2$ Wess-Zumino model is ultraviolet finite
and that the index is given by $p-1$ \cite{Jaffe:1987qz}. This seems to
be in contradiction with our result, as the $(p-1)^N$ zero modes exist for
all finite lattices and, by the same arguments as for the $\mN=1$ model,
remain zero modes in the continuum limit.

We suggest the following solution for this problem. Remember that
our lattice Hamiltonian $H$ contains not only the discretized version of the
continuum Hamiltonian $P_0$ but also the central charge $\mZ$, i.e.
\eqnl{
H = P_0 + \mZ.}{n2weak1}
Furthermore, both $P_0$ and $\mZ$ contain the
lattice derivative which couples fields at different
lattice sites. If we choose in the
strong-coupling limit a zero mode that varies from lattice
point to lattice point, both $P_0$
and $\mZ$ may become very large but will, nevertheless, add up to zero.
In the continuum limit the energy $P_0$ may be infinite in which case
this rapidly varying zero mode is only a lattice artifact. On the other
hand, if we choose the same zero mode for each lattice site, then $P_0$
as well as $\mZ$ should be zero in the continuum limit. Thus, 
there are exactly $p-1$ such modes. We are planning to test this
proposal in a perturbative calculation of $P_0$ and $\mZ$.
The results will be presented elsewhere.

\section{Conclusions}

In this paper we have related 
Dirac operators defining supersymmetric quantum
mechanical systems in high-dimensional spaces
\cite{Kirchberg:2004za} to Wess-Zumino models on a spatial
lattice in $1+1$ dimensions.
After a very particular dimensional reduction the
square of $\ui\di$ can be identified with
the super-Hamiltonians of latticized Wess-Zumino
models. This way we discovered a natural connection between
discretized supersymmetric field theories
and supersymmetric quantum mechanics.

We have recalled
the continuum formulation of Wess-Zumino models and
discussed their lattice versions. For the
case of simple ($\mN=1$) and extended
($\mN=2$) supersymmetry, we have derived the
corresponding Dirac operators. Furthermore,
all ground states for the free massive models and the
interacting theories in the limit of strong coupling
have been constructed.

Different realizations of lattice derivatives
have been discussed and their properties -- in
particular from the point of view of supersymmetric
quantum mechanics -- have been analyzed.
Our results on the
number of zero modes do not depend on the
particular lattice derivative, as long as some mild
assumptions are fulfilled.

Employing powerful theorems from functional analysis
we were able to relate the strong and weak
coupling regions. For $\mN=1$ it turns out
that generically the number
of zero modes is determined by the strong-coupling limit.
There is a single (no) zero mode, if the degree
of the superpotential is even (odd). For $\mN=2$
we find at least $(p-1)^N$ zero modes, where
$p$ is the degree of the superpotential and $N$ the
number of lattice sites.  This number is far off
the correct continuum result, which predicts $p-1$
zero modes, a serious problem which has been observed
earlier in \cite{Elitzur:1983nj}.

We have explained this paradox as follows:
the lattice Hamiltonian $H$ does not only contain
the continuum Hamiltonian $P_0$ but also
additional terms which (for antisymmetric
lattice derivatives) are to be interpreted as a lattice
version of the central charge $\mathcal{Z}$.
On the lattice, $P_0$ and $\mathcal{Z}$
cancel pairwise for the huge number of zero modes
under discussion, even though neither $P_0$ nor
$\mZ$ is zero in the continuum limit, except for exactly $p-1$
of the modes. 

Our Dirac operators clearly deserve further studies. For
instance, the application of (standard) index theorems
to the case at hand should reveal new information
about the field theories. We also plan to
extend our results to Dirac operators on curved
manifolds, which can be reinterpreted as nonlinear
$\sigma$-models from the field theory point of view.

\section*{Acknowledgements}
We thank Hans Triebel for very useful discussions on
aspects of functional analysis and Falk Bruckmann
and Ivo Sachs for useful conversations.
A.K. acknowledges 
support by the Studienstiftung des Deutschen Volkes.

\begin{appendix}
\section{The SLAC Operator}
In this appendix we introduce and discuss the
nonlocal SLAC derivative. It can be used to define
chiral fermions without fermion-doubling.

First we consider \emph{real valued} scalar fields on the spatial
lattice. They maybe interpreted as wave
functions of a quantum mechanical system with
Hilbert space $\R^N$, equipped with
the scalar product
\eqnn{(\phi,\chi)=\sum_{n=1}^N \bar\phi(n)\chi(n).}
Although  the fields are real it is useful to
embed them in the space of complex valued
lattice fields.
For a normalised function we interpret
$\vert\phi(n)\vert^2$ as probability
for finding the 'particle' at the lattice site
$n$. Expectation values of functions of
the 'position' operator $\hat n$ are
\eqnl{
\la f(\hat n)\ra_\phi=\sum \bar\phi(n)f(n)\phi(n).}{qmcom3}
We want to derive a similar formula for
expectation values of functions of the
momentum operator. For that aim we Fourier
transform the wave function as follows
\eqnl{
\tilde\phi(p_k)=\frac{1}{\sqrt{N}}\sum_{n=-N'}^{N'}
\ue^{-\ui p_k\,n}\phi(n),\mtxt{where}N'=\ft12(N\ms 1),\quad
p_k=\frac{2\pi}{N}k.}{qmcom5}
The inverse Fourier transformation reads
\eqnl{
\phi(n)=\frac{1}{\sqrt{N}}\sum_{k=-N'}^{N'} \ue^{\ui p_k\,n}\tilde
\phi(p_k),\quad n=-N',\dots,N'.}{qmcom7}
We have chosen the symmetric summation to end up with
a real difference operator.
For periodic fields $n$ must be integer
and this is the case for space lattices
with an \emph{odd number of sites}.
For a normalized $\phi$ the Fourier transform $\tilde \phi$
is normalized as well and we may interpret
$\vert\tilde\phi(p_k)\vert^2$ as probability for
finding the 'momentum' $p_k$. With this interpretation
we obtain
\eqnl{
\la f(\hat p)\ra_\phi=
\sum_{k=-N'}^{N'} f(p_k)\,\vert\tilde\phi (p_k)\vert^2
=\sum_{nn'}\bar\phi_n f(\hat p)_{nn'}\phi_{n'}}{qmcom8}
with matrix elements
\eqnn{
f(\hat p)_{nn'}=\frac{1}{N}\sum_{k=-N'}^{N'} \ue^{\ui p_k(n-n')}f(p_k).}
With the help of the generating function
\eqnl{
Z(x)=\sum_{k=-N'}^{N'} \ue^{\ui p_k x}=\frac{\sin (\pi x)}{\sin (\pi x/N)},}{qmcom10}
we can calculate all matrix elements of $f(\hat p)$.
Now we are ready to define the real, nonlocal and antisymmetric
lattice operator $\pSL=\ui\hat p$.
The matrix elements are
\eqnl{
f(\pSL)_{nn'}=\frac{1}{N} f\left(\frac{\ud}{\ud x}\right)
Z(x)\Big\vert_{x=n-n'}\;.}{qmcom12}
As expected $\pSL$ is a Toeplitz matrix
with elements
\eqnl{
(\pSL)_{nn'}=(-)^{n-n'}\frac{\pi/N}{\sin\big(\pi(n-n')/N\big)},\quad
\mtxt{for}n\neq n',}{qmcom13}
and the elements on the diagonal vanish,
$(\pSL)_{nn}=0$.

\section{Analyticity of Perturbation}
\label{inequ}

In the following we consider operators on the Hilbert space
\eqnl{
\mH = \text{L}_2(\R^d, \ud^d x) \otimes \C^D}{inequ1}
for $D\in \N$ with norm
\eqnl{
\| f \|^2 = \sum_{i=1}^D \| f_i \|_{\text{L}_2}^2\, ,\quad
f = (f_1,\ldots,f_D) \in \mH.}{inequ2}
Here, $\| \cdot \|_{\text{L}_2}$ denotes the familiar
$\text{L}_2$-norm.

\subsection{The $\mN=1$ Case}
\label{n1inequ}

For the Wess-Zumino model on the lattice
with $\mN=1$ supersymmetry we take $D=2^N$, $d=N$ ($N=$ number of lattice points)
and consider the (unperturbed) operator \refs{reda10}
\eqnl{
A_0 = \sum_{n=1}^N \left( - \ui
\psi_1(n) \p_n + \psi_2(n) W'(x_n) \right).}{n1inequ1}
We recall that $W$ is a polynomial of degree $\text{deg}(W)=p>1$ and
$\psi_\alpha(n)$ are Hermitian
$D\times D-$matrices obeying the Clifford algebra
\eqnl{
\{ \psi_\alpha(n) , \psi_\beta(n^\prime) \} = 2 \delta_{\alpha \beta}
\delta(n,n^\prime), 
\quad \alpha,\beta=1,2, \quad 
n,n^\prime=1,2,\ldots,N.}{n1inequ2}
The operator $A_0$ with domain of definition
\eqnn{
\mD(A_0) = C^\infty_{\text{c}}(\R^N) \otimes \C^D}
is essentially self-adjoint, where
we write $C^\infty_{\text{c}}(\R^N)$ for the
$C^\infty$-functions with compact support in $\R^N$. A simple
calculation using \refs{n1inequ2} shows
\eqnn{
(A_0)^2 = \sum_n \left( - \p_n \p_n + W'(x_n) W'(x_n) - \ui \psi_1(n)
\psi_2(n) W''(x_n) \right).}

\textbf{Closure of the Operator $A_0$}

To determine the closure $\bar{A}_0$ of the operator $A_0$
we have to find the closure of its domain
$\mD(A_0)$ with respect to the
norm
\eqnl{
\| f \|^2_{A_0,a} = a \| f \|^2 + \| A_0 f \|^2, \quad a>0.}{abbrev1}
Note that
these norms are equivalent for all $a>0$. Using the abbreviation
\eqnl{\rho_p = 1 + |x|^{p-1},}{abbrev2}
we can prove the following

\textbf{Lemma:} There exist constants
$a,b_1, b_2 >0$ such that
\eqnl{
\| f' \|^2 + b_1 \| \rho_p f \|^2 \leq
\| f \|^2_{A_0,a} \leq \| f' \|^2 + b_2 \|
\rho_p f \|^2}{closure1}
holds for all $f \in \mD(A_0)$.

In the Lemma we used the short hand notation
$\| f' \|^2 = \sum_m \| \p_m f \|^2$.

\emph{Proof:}
First, we show that only the degree $\text{deg}(W)=p$ is important for
terms like $\sum_n \| W'(x_n) f \|^2$ with $f \in \mD(A_0)$.
Indeed, we find
\eqnl{
\sum_n \| W'(x_n) f \|^2 \leq N  a_1 \|\rho_p f \|^2, \quad a_1 = 
\left\| \frac{W'(x_n)}{\rho_p} \right\|^2_\infty,}{closure2}
where $\| \cdot \|_\infty$ denotes the supremum norm. The factor $N$
arises from the sum over $n$, as $a_1$ does not depend on $n$.
Similar we obtain
\eqngrl{
\| \rho_p f \|^2&=&\left\| \frac{\rho_p} {\sqrt{1 + \sum_n
W'^2(x_n)}}\sqrt{1 + \sum W'^2(x_n)}
\,f \right\|^2}
{&\leq& a_2 \left(\| f \|^2 + \sum \| W'(x_n) f
\|^2\right), \quad
a_2 = \left\| \frac{\rho_p}
{\sqrt{1+\sum W'^2(x_n)}} \right\|^2_\infty.}{closure3}
Now, it is easy to prove the second inequality in \refs{closure1},
\eqngrl{
a \| f \|^2 + \| A_0 f \|^2&\overset{\refs{closure2}}{\leq}&
\| f' \|^2 + a \| f \|^2 + N a_1
\|\rho_p f \|^2 + \sum_n \| f\| \; \| W''(x_n) f \|}
{&\leq& \| f' \|^2 + \underbrace{(a + N a_1 + N a_3)}_{b_2}
\|\rho_p f\|^2,}
{closure5}
with $a_3 =\left\|\frac{W''(x_n)}{\rho_p}\right\|_\infty$. 
We used that the matrix-norm of $\psi_\alpha(n)$ is
one, since its eigenvalues are $\pm 1$. In the last
inequality we made use of $\| f \| \leq \|\, \rho_p f \|$
which holds for all $f \in \mD(A_0)$.

The other inequality in \refs{closure1} is
more difficult to prove. With \refs{closure3} we get
\eqnl{
a \| f \|^2 + \| A_0 f \|^2 \geq \| f' \|^2 + \frac{1}{a_2} \|
\rho_p f \|^2+(a - 1) \| f \|^2 - \sum \| f \| \; \|
W''(x_n) f \|.}{closure6}
In order to obtain a positive constant $b_1$ in our
lemma we must be rather careful with our estimates for
the last term in \refs{closure6}.
We introduce a ball of radius $R$ and split $f$ into
two parts, $f = f_< + f_>$, where $f_<$ has its support
inside the ball $f_>$ outside the ball. We obtain
\eqnl{
\sum_n \| f \| \; \| W''(x_n) f \| = \sum \left( \| f_< \| \; \|
W''(x_n) f_< \| + \| f_> \| \; \| W''(x_n) f_> \| \right),}{closure7}
where the terms containing both $f_<$ and $f_>$ vanishes. Let
us now consider the two terms separately. First, we obtain
\eqnl{
\sum \| f_> \| \; \| W''(x_n) f_> \| \leq
N a_4(R) \| \rho_p f \|^2, \quad a_4(R) = \left\|
\frac{W''(x_n)} {\rho_p} \right\|_{\infty,>},}{closure8}
where we have introduced the supremum norm
$\| \cdot \|_{\infty,>} = \text{sup}_{|x|>R} \{ | \cdot | \}$. For
large $R$ we have $a_4(R) \sim 1/R$ such that
$a_4$ gets arbitrarily small for big balls. Second,
we obtain
\eqnl{
\sum \| f_< \| \; \| W''(x_n) f_< \| \leq N a_5(R) \| f \|^2, \quad
a_5(R) = \| W''(x_n) \|_{\infty,<},}{closure9}
with
$\| \cdot \|_{\infty,<} = \text{sup}_{|x|<R} \{ | \cdot | \}$. For
$R \rightarrow \infty$ we have $a_5(R) \rightarrow \infty$. Altogether,
we find
\eqnl{
a \| f \|^2 + \| A_0 f \|^2 \geq \| f' \|^2 +
\left(a-1-N a_5(R)\right)
\| f \|^2 + \underbrace{(1/a_2 - N a_4(R))}_{b_1} \|
\rho_p f\|^2.}{closure10}
In a first step we choose $R$ large enough such that
$b_1$ is positive. In a second step we choose $a$ large
such that the constant in front of $\| f \|^2$ is positive
as well. This finishes the proof of our lemma. 

Since all norms
\eqnl{
\| f \|^2_b \equiv \| f'\|^2 + b\, \| \rho_p f \|^2}
{closure11}
are equivalent for $b>0$, the Lemma implies that these
norms are equivalent to the norms $\| f \|^2_{A_0,a}$
in \refs{abbrev1}.
Therefore, the closure of $\mD(A_0)$ with respect to the norm
\refs{abbrev1} coincides with the closure with respect to
$\| \cdot \|_b$. This closure is given by
\eqnl{
\mD(\bar{A_0}) = \left\{ f \in W_2^1(\R^N) \otimes \C^D :
\| \rho_p f \| < \infty \right\} \equiv 
W_2^1(\R^N,\rho_p^2) \otimes \C^D.}{closure13}
Here $W_2^1(\R^N)$ is the Sobolev space with first weak-derivative in
$\text{L}_2$.

\textbf{Perturbation}\\
Let us perturb the operator $A_0$ by the operator $A_1$ in \refs{reda10},
\eqnl{
A_1 = - \sum_{m,n=1}^N x_m \left(\p\right)_{mn} \psi_2(n).}{perturb1}
The operator $A_1$ is self-adjoint with
$\mD(A_1)=\text{L}_2(\R^N,\tilde{\rho}) \otimes
\C^D \supset \mD(\bar{A_0})$, with
$\tilde{\rho}$-weighted Lebesgue measure, where
$\tilde{\rho}(x) = (1+|x|)^2$.
From the following Lemma we will derive useful information
about the nature of the perturbation.

\textbf{Lemma:} For all $\lambda \in \R$ and \emph{arbitrarily small}
$\epsilon>0$ there exists a $C_\epsilon>0$ such that
\eqnl{
\| \lambda A_1 f \| \leq \epsilon \| A_0 f \| + C_\epsilon \| f \|,
\quad \forall f \in \mD(\bar{A_0}).}{perturb2}

\emph{Proof:} We prove the inequality for all $f \in \mD(A_0)$.
Then it holds for all elements in the closure as well. As before
we split $f=f_< + f_>$. First, we note
\eqnl{
\| \lambda A_1 f_< \| \leq |\lambda| N^2 a(R)\, \| f \|, \quad a(R) = \|
x_n \|_{\infty,<}
\cdot \text{max}\{ |\p_{mn}| : m,n=1,\ldots,N\}.}{perturb4}
For $R \rightarrow \infty$ the constant $a(R)$ tends to infinity.
Next, we have
\eqngrl{
\| \lambda A_1 f_> \| &\overset{\refs{closure1}}{\leq}& |\lambda|
N^2 b(R) \left(c\, \| f \| +\| A_0 f \|\right),}
{b(R)&=& \left\|\frac{x_n} {\rho_p(x)}
\right\|_{\infty,>} \cdot \text{max}\{ |\p_{mn} | : m,n=1,\ldots,N\}}{perturb5}
for some positive constant $c$. For big $R$ the
constant $b(R)$ tends to zero. We choose the ball
big enough such that $|\lambda| N^2 b(R) = \epsilon$ and set
$C_\epsilon= \epsilon c + |\lambda|N^2 a(R)$. 
Note that the latter constant may become huge.

\textbf{Self-adjointness}\\
We did prove that $\bar{A}_0$ is a self-adjoint operator. Clearly
$\lambda A_1$ is symmetric on $\mD(\bar{A}_0)$.
Furthermore, \refs{perturb2} shows that
$\lambda A_1$ is $\bar{A}_0$-bounded with relative bound
less than one. The famous Kato-Rellich Theorem,
see Theorem X.12 in \cite{Reed}, states
that under these conditions the operator
\eqnl{
Q_1(\lambda) = A_0 + \lambda A_1}{perturb6}
is self-adjoint with domain $\mD(\bar{A}_0)$.
We conclude that $Q_1(\lambda)$ is a family of
self-adjoint operators with common domain of definition
$\mD(\bar{A}_0)$.

\textbf{Analyticity of Eigenvalues}\\
In the following we prove that $Q_1(\lambda)$ is an
\emph{analytic family} in the sense of Kato
for all real $\lambda$. We have seen that $Q_1(\lambda)$
is self-adjoint for real $\lambda$. For a
self-adjoint and analytic family it is known that
the eigenvalues depend analytically on the parameter $\lambda$, see for
example Theorem XII.13 in \cite{Reed}.

For an arbitrary real $\lambda_0$ the perturbation
$\lambda_0 A_1$ is $\bar{A}_0$-bounded
with arbitrary small relative bound (\ref{perturb2}).
Then, it is easy to see that $A_1$ is $Q_1(\lambda_0)$-bounded.
It follows that for small $\epsilon$ the
operators $Q_1(\lambda_0+\epsilon)$ form an analytic family of
type (A) \cite{Reed} and therefore also an analytic family in the sense
of Kato. But as $\lambda_0 \in \R$ is arbitrary, we haven proved that
$Q_1(\lambda)$ is analytic for all real $\lambda$.

Actually, the cited Theorem XII.13 \cite{Reed}
above is only valid for isolated
eigenvalues with finite degeneracy or equivalently
for eigenvalue in the discrete spectrum.
In the following we prove that the spectrum of
$Q_1(\lambda)$ is discrete by proving this statement
for its square, $H(\lambda) = Q_1(\lambda)^2$. $H(\lambda)$ is
self-adjoint with domain of definition given by
\eqngrl{
\mD\left(H(\lambda)\right) &\equiv& \left\{ f \in \mD(\bar{A}_0) :
Q_1(\lambda) f \in \mD(\bar{A}_0) \right\} }
{&=&W_2^2(\R^N,\rho') \otimes \C^D, \qquad
\rho'(x)= \left(1+|x|^{2p-2}\right)^2}{defi}
and it is semibounded
\eqnl{
H(\lambda) \geq 0.}{posi}
Such operators possess entirely discrete spectra if
and only if its resolvent is a compact operator, see Theorem XIII.64 in
\cite{Reed}. In the following we prove that
$H(\lambda)$ has compact resolvent for all $\lambda \in \R$.

We must prove that the image of a bounded subset of the Hilbert space,
say
\eqnl{
\{f \in \mH:\|f\| <1\},}{bound1}
is mapped to a precompact set under the map $(H-z)^{-1}$
for some $z$ in the resolvent of $H$. The image is given by
\eqnl{
\{ g \in \mD(H) : \| (H-z) g \| < 1\}.}{bound2}
As earlier we split $g$ into $g_>$
and $g_<$ and obtain for large enough radii $R$
the inequality
$\| g \| \leq \| g_< \| + \epsilon$. For a compact ball
$\mathcal{B} = \{ x\in  \R^N: |x|\leq R\}$ we have
Sobolev's embedding theorem and there is an $\epsilon$-net
$g_j \in W_2^2(K,\rho')$, $j=1,\ldots,N_\epsilon$ with
$\| g_< - g_j \| < \epsilon$ for one $j\in\{1,\ldots,N_\epsilon\}$.
We extend the $g_j$ by zero to the region outside
the ball and obtain
\eqnl{
\| g-g_j\| &\leq 2 \epsilon}{bound3}
for any $g$ in the image of the unit ball under $(H-z)^{-1}$ and a specific
$j \in \{ 1, \ldots,N_\epsilon\}$. We conclude that there is a
$2\epsilon$-net of the image and therefore the image is precompact.
This completes our proof.

\textbf{Stability of the Index}\\
We have shown that the eigenvalues are analytic functions of the
parameter $\lambda$ on the whole real axis. It follows at onces that
the index -- the difference of bosonic zero modes and fermionic zero
modes -- is also an analytic function and, as
the index only takes on integer values, is constant.

An alternative and elegant proof of this statement
can be given with the help of the theorem that
a \emph{relatively compact perturbation} does not change
the index \cite{Kato}. Indeed, inequality (\ref{perturb2})
implies that our perturbation is relatively
compact\footnote{We thank H.~Triebel for the proof
of this statement.}.

\subsection{The $\mN=2$ Case}
\label{n2inequ}
After the detailed investigation of the $\mN=1$ case
we shorten our discussion for $\mN=2$. In
what follows we consider the real part of
the complex supercharge in \refs{redb42}
\eqnl{
B_0 = \sum_{n=1}^{N}  \left( -\ui \psi_1^1(n) \p_{x_n} - \ui \psi_1^2(n)
\p_{y_n} + \sum_{a=1}^2 \psi_2^a(n) W_{,a}(x_n,y_n) \right)}{n2inequ1}
in the strong-coupling limit. For $\mN=2$ supersymmetry
the $\psi_\alpha^a$ are $D$-dimensional Hermitian matrices
 obeying the Clifford algebra, with $D=2^{2N}$.
The function $W(x,y)$ is harmonic and and thus
is the real part of an analytic function $F(x+\ui y)$.
As in chapter \ref{chapwzm} we use the notation
$W,_1(x,y) = \p_x W(x,y)$ and $W,_2(x,y) = \p_y W(x,y)$.
As domain of definition we take
\eqnl{
\mD(A_0) = C^\infty_{\text{c}}(\R^{2N}) \otimes \C^D,\qquad
D=2^{2N},}{n2inequ2}
such that $B_0$ is essentially self-adjoint. We introduce
the potential
\eqnl{
K(x,y) = \sum_a W,_a^2(x,y).}{n2inequ4}
For large radii $r$ only the leading power of $W$ is relevant.
Therefore, we may consider the particular case
\eqnl{
W(x,y) = \frac{\kappa}{p} \Re z^p}{pot}
for which we find $K(x,y) = \kappa r^{p-1}\rightarrow \infty$
in all directions for $r \rightarrow \infty$.

The perturbation contains the lattice derivative,
\eqnn{
B_1 =\sum_{m,n=1}^{2N} \left( x_m (\p)_{mn} \psi_2^2(n) +
y_m (\p^\dagger)_{mn} \psi_2^1(n) \right).}
Replacing in the estimates for the case with $\mN=1$
supersymmetry the potential $W'(x_n)$ by $K(x_n,y_n)$
leads to analogous results in the $\mN=2$ case.
Again all eigenvalues are analytic functions of the parameter
$\lambda$ and in particular the index does not depend on this parameter.
\end{appendix}

\providecommand{\href}[2]{#2}\begingroup\raggedright\endgroup

\end{document}